\documentclass[11pt]{article}
\pdfoutput=1
\usepackage{jheppub}

\usepackage{enumerate}
\usepackage{amsmath,amssymb,epsfig,cancel}
\usepackage{color}
\usepackage{bbold}
\usepackage{graphicx}
\usepackage{physics}
\usepackage{slashed}
\usepackage{caption}
\usepackage{braket}
\usepackage[english]{babel}

\definecolor{mypink}{RGB}{219, 48, 122}

\usepackage{caption}
\usepackage{subcaption}

\title{Holography for 2d $\mathcal{N}=(0,4)$ quantum field theory}

\author{Kostas Filippas}

\affiliation{Department of Physics, Swansea University, Swansea SA2 8PP, United Kingdom}

\emailAdd{kphilippas@hotmail.com} 

\abstract{We study the correspondence between AdS$_3$ massive IIA supergravity vacua and two-dimensional $\mathcal{N}=(0,4)$ quiver quantum field theories. After categorizing all kinds of gravity solutions, we demystify the ones that seem to reflect anomalous gauge theories. In particular, we prove that there are bound states of D-branes on the boundary of the space which provide the dual quiver theory with exactly the correct amount of flavor symmetry in order to cancel its gauge anomalies. Then we propose that the structure of the field theory should be complemented with additional bifundamental matter, which we argue may only be $\mathcal{N}=(4,4)$ hypermultiplets. Finally, we construct a BPS string configuration and use the old and new supersymmetric matter to build its dual ultraviolet operator. During this holographic synthesis, we uncover some interesting features of the quiver superpotential and associate the proposed operator with the same classical mass of its dual BPS string.
\\[5pt]
 }
 
\keywords{AdS$_3$, two dimensions, superfields, quivers, D-branes, bound states.\\[30pt]}

\subheader{}
\vspace{3cm}
\dedicated{Dedicated to the memory of Muhammad Al-Arab and Muhammad Gulzar.}


\begin{document}
\def\Tr{{\textrm{Tr}}}

%%%%%%%%%%%%%%%%%%%%%%%%%%%%%%%%%%%%%%%%%%%%%%%%%%%%%%%%%%%%%%%%

%\hfill { CERN-TH-}

%%%%%%%%%%%%%%%%%%%%%%%%%%%%%%%%%%%%%%%%%%%%%%%%%%%%%%%%%%%%%%%%

\maketitle

\section{Introduction}
The AdS/CFT correspondence constitutes a primo realization of the holographic principle while it ties string theory to the most well-studied particle theories we possess. In other words, besides being a conceptual breakthrough on its own right, holography brings strong confidence that a complete quantum theory of gravity shines upon the physics of the superstring. Nonetheless, the power of this duality does not limit itself in supporting quantum gravity but also unravels the properties of certain supersymmetric quantum field theories that otherwise are yet out of our reach through the standard methods or techniques.

While over the years many type II supergravity solutions have made their appearance in the holographic arena, there is a certain kind that has recently been poping up more frequently and has become quite popular. These are supergravity backgrounds whose entirety of fields is defined by functions of the coordinates of the internal manifolds and are dual to supersymmetric quiver gauge theories. Studying those backgrounds ultimately boils down to understanding their defining functions. The dual physics of these vacua is generally described by supersymmetric conformal field theories (SCFTs), which for $d<4$ are assumed to be strongly coupled IR fixed points that flow to better-understood ultraviolet (UV) quiver field theories through the renormalization group (RG) equations. The latter are defined on supersymmetric multiplets of fundamental fields, whose interactions are usually well-defined and provide an understandable particle theory.

SCFTs exist exclusively in $d<7$ dimensions \cite{Nahm:1977tg} and there has been intensive work on all of their diversity, both field theoretically and holographically. In six dimensions, an infinite family of $\mathcal{N}=(0,1)$ theories has been discussed in \cite{Brunner:1997gf,Hanany:1997gh,Gaiotto:2014lca,Cremonesi:2015bld,Apruzzi:2013yva,Apruzzi:2014qva,Apruzzi:2015wna,Apruzzi:2017nck,Passias:2015gya,Bobev:2016phc,Macpherson:2016xwk,Filippas:2019puw}. In five dimensions, solutions in a variety of supersymmetry were analyzed in \cite{Lozano:2012au,DHoker:2016ujz,DHoker:2016ysh,DHoker:2017mds,Gutperle:2018vdd,Fluder:2018chf,Bergman:2018hin,Lozano:2018pcp}. For $\mathcal{N}=2$ supersymmetry in four dimensions there has been a fruitful study in \cite{Gaiotto:2009we,ReidEdwards:2010qs,Aharony:2012tz,Bah:2015fwa,Nunez:2018qcj,Nunez:2019gbg,Bah:2019jts}, while three dimensional $\mathcal{N}=4$ theories were discussed in \cite{Gaiotto:2008ak,DHoker:2007hhe,DHoker:2008lup,Assel:2011xz,Lozano:2016wrs}.

The case of AdS$_3$ supergravity solutions is somewhat unique. Three dimensional gravity as well as the algebra of two dimensional field theory make the study of AdS$_3$ holography of particular interest and this is reflected on the rich literature regarding the subject, some representatives of which are \cite{Witten:1997yu,Seiberg:1999xz,Kutasov:1998zh,Maldacena:1997de,Martelli:2003ki,Castro:2008ne,Couzens:2019wls,Kim:2015gha,Gadde:2015tra,Lawrie:2016axq,Couzens:2017way,Lozano:2015bra,Kelekci:2016uqv,Macpherson:2018mif,Legramandi:2019xqd,Lozano:2019emq}.

Another family of such AdS$_3$ solutions was recently introduced in \cite{Lozano:2019emq,Lozano:2019jza,Lozano:2019zvg,Lozano:2019ywa}. These massive IIA vacua are associated with D2-D4-D6-D8 Hanany-Witten brane set-ups \cite{Hanany:1996ie} and were first build in \cite{Lozano:2019emq}. The D2 and D6-branes exist as fluxes and they are dual to gauge symmetries, while the D4 and D8-branes live explicitly in the background and provide dual flavor symmetries. In \cite{Lozano:2019zvg} a particular class of them that exhibits the local geometry AdS$_3\times$S$^2\times$CY$_2\times\mathbb{R}$ was distinguished and was proposed to be dual to two-dimensional quiver quantum field theories with $\mathcal{N}=(0,4)$ supersymmetry. Some holographic aspects of these quivers were studied in \cite{Speziali:2019uzn,Lozano:2020bxo}. Those are the theories that we are about to consider.

The defining functions of a supergravity solution render the form of the fields on the gravity side of the correspondence, while they accordingly shape the exact structure of the dual quiver field theory. In order to validate the correspondence and study the whole range of its potential, one should explore the various properties of these functions and confirm that every single time they make perfect sense on their dual field-theoretical attribution. This makes up the starting point of this article, where we take the most unusual choice of such defining functions which seems to give an anomalous dual quantum field theory. By carefully focusing on the right regions of the supergravity background we discover D-branes that are realized as global symmetries in the dual quiver structure, providing exactly the flavors needed to cancel the apparent gauge anomalies. Due to strong Ramond-Ramond (RR) fluxes on the boundary of the space these D-branes come exclusively in bound states, forming polarizations that provide flavor symmetries in an idiosyncratic way.

Observing the quiver structure of the theories under consideration, we realize that there must be some linking multiplets missing. Such multiplets bind color D2 with flavor D4-branes and color D6 with flavor D8-branes, while it is shown that those may only be $\mathcal{N}=(4,4)$ hypermultiplets corresponding to suspended superstrings between D2 and D4-branes or D6 and D8-branes in the ancestral Hanany-Witten set-up.

The existence of this new matter complements the quiver structure, while it seems to be also vital in the construction of the dual operator for a particular BPS string state. To be precise, after picking a semiclassical string configuration connecting two stacks of D-branes in the background, we prove that this is a BPS state and propose a string of scalar fields as its dual UV operator. We argue that this is a unique choice of a dual operator and, while two-dimensional scalars have mass dimension zero implying a vanishing conformal dimension for that operator, we conclude that the latter property is attained non-perturbatively. That is, we bring to the surface the superpotential of the UV quiver theory to find interactions between the scalars inside the operator, supporting the idea of a totally non-perturbative anomalous dimension at the IR of the RG flow.

Finally, we find that scalars inside the vector superfields should obtain a vacuum expectation value (VEV) through a Fayet-Iliopoulos term due to the U$(1)$ theory inside each U$(N)$ gauge group. Superpotential interactions between the vector and hypermultiplets then dictate that bifundamental matter acquires a mass, ultimately associating the dual UV operator with a classical mass equal to that of the BPS string. Since the operator mass is a sum of all the individual scalar field masses, this renders the operator very much alike to a classical bound state of particles dual to a bound string state between D-branes.

The plan of this paper is as follows. In Section \ref{Section2} we review the massive IIA supergravity backgrounds and quantum field theory first constructed in \cite{Lozano:2019emq}. We also give a brief but complete summary of two-dimensional $\mathcal{N}=(0,4)$ quantum field theory that is useful in understanding gauge anomalies, R-current charges and superpotentials between multiplets, all basic ingredients for the self-containment of the present work. In Section \ref{Section3} we study special solutions of vacua that naively give anomalous quiver theories and show how these are canceled by flavor symmetries produced by dielectric branes on the boundary of the space. In Section \ref{Section4} we illustrate that new matter should be added in the structure of the field theory in the form of $\mathcal{N}=(4,4)$ hypermultiplets. Finally, in Section \ref{Section5} we construct a BPS string soliton and propose a dual operator, which both seem to exhibit the same classical mass.

\section{AdS$_3$ massive IIA vacua vs $\mathcal{N}=(0,4)$ theory}\label{Section2}
\subsection{The supergravity solutions}
In \cite{Lozano:2019emq} a new family of AdS$_3$ massive IIA supergravity solutions with $\mathcal{N}=(0,4)$ supersymmetry  was introduced. A subclass of these solutions with local geometry AdS$_3\times$S$^2\times$CY$_2\times$I$_\rho$ was conjectured in \cite{Lozano:2019jza,Lozano:2019zvg,Lozano:2019ywa} to be dual to $\mathcal{N}=(0,4)$ quiver quantum field theories in two dimensions. These vacua have an NS NS sector, in string frame,

\begin{equation}
\begin{split}
\dd s^2\:=\:\frac{u}{\sqrt{h_4h_8}}\left(\,\dd s^2_{\mbox{\tiny AdS$_3$}}+\frac{h_4h_8}{4h_4h_8+(u')^2}\,\dd s^2_{\mbox{\tiny S$^2$}}\right)+\frac{\sqrt{h_4h_8}}{u}\dd\rho^2+\sqrt{\frac{h_4}{h_8}}\,\dd s^2_{\mbox{\tiny CY$_2$}}\:,\hspace{2cm}\\[10pt]
B_2\:=\:\frac{1}{2}\left(2k\pi-\rho+\frac{uu'}{4h_4h_8+(u')^2}\right)\,\mbox{vol(S$^2$)}\:,\hspace{2cm}e^{-\phi}\:=\:\frac{h_8^{\frac{3}{4}}}{2h_4^{\frac{1}{4}}\sqrt{u}}\sqrt{4h_4h_8+(u')^2}\,,\label{GeneralAdS3background}
\end{split}
\end{equation}
where $u,h_4,h_8$ are functions of the coordinate $\rho$, defining this family of supergravity backgrounds. Note that we also allow for large gauge transformations $B_2\rightarrow B_2+\pi k\,\mbox{vol}_{S^2}$, every time we cross a $\rho$-interval $[2\pi k,2\pi(k+1)]$, for $k=0,...,P$. The RR sector reads
\begin{equation}
\begin{split}
\hat{F}_0\;&=\;h_8'\:,\hspace{1.3cm}\hat{F}_2\;=\;-\frac{1}{2}\left(h_8-h_8'(\rho-2\alpha'\pi k)\right)\mbox{vol(S$^2$)}\:,\\[10pt]
\hat{F}_4\;&=\;\left(\partial_\rho\left(\frac{uu'}{2h_4}\right)+2h_8\right)\dd\rho\wedge\mbox{vol(AdS$_3$})-h'_4\:\mbox{vol(CY$_2$)}\,,
\end{split}\label{RRsector}
\end{equation}\\
where $\hat{F}=e^{-B_2}\wedge F$ is the Page flux. These functions are locally constrained as

\begin{equation}
h_4''\;=\;h_8''\;=\;u''=0\,,\label{SUGRAeom}
\end{equation}
where the first two equations come from the Bianchi identities, while the last comes from supersymmetry. This results in piecewise linear functions
\begin{equation}
h_4(\rho)\:=\:\left\lbrace\begin{array}{ccc}
\alpha_0+\frac{\beta_0}{2\pi}\rho &\hspace{1cm}0\leq\rho\leq2\pi &\\
\alpha_k+\frac{\beta_k}{2\pi}(\rho-2\pi k) &\hspace{1cm}2\pi k\leq\rho\leq2\pi(k+1) &k=1,...,P-1\;,\\
\alpha_P+\frac{\beta_P}{2\pi}(\rho-2\pi P) &\hspace{1cm}2\pi P\leq\rho\leq2\pi(P+1) &
\end{array}\right.\label{h4piecewise}
\end{equation}
\begin{equation}
h_8(\rho)\:=\:\left\lbrace\begin{array}{ccc}
\mu_0+\frac{\nu_0}{2\pi}\rho &\hspace{1cm}0\leq\rho\leq2\pi &\\
\mu_k+\frac{\nu_k}{2\pi}(\rho-2\pi k) &\hspace{1cm}2\pi k\leq\rho\leq2\pi(k+1) &k=1,...,P-1\;,\\
\mu_P+\frac{\nu_P}{2\pi}(\rho-2\pi P) &\hspace{1cm}2\pi P\leq\rho\leq2\pi(P+1) &
\end{array}\right.\label{h8piecewise}
\end{equation}\\
while $u=a+b\rho$ globally, for supersymmetry to be preserved. Note that $P,\alpha_k,\mu_k$ have to be large for the supergravity limit to be trusted, while continuity of these equations along $\rho$ implies $\mu_k=\sum_i^{k-1}\nu_j$ and $\alpha_k=\sum_i^{k-1}\beta_j$.

\begin{figure}[t!]
    \centering
    %\subfloat[label 1]
    {{\includegraphics[width=8cm]{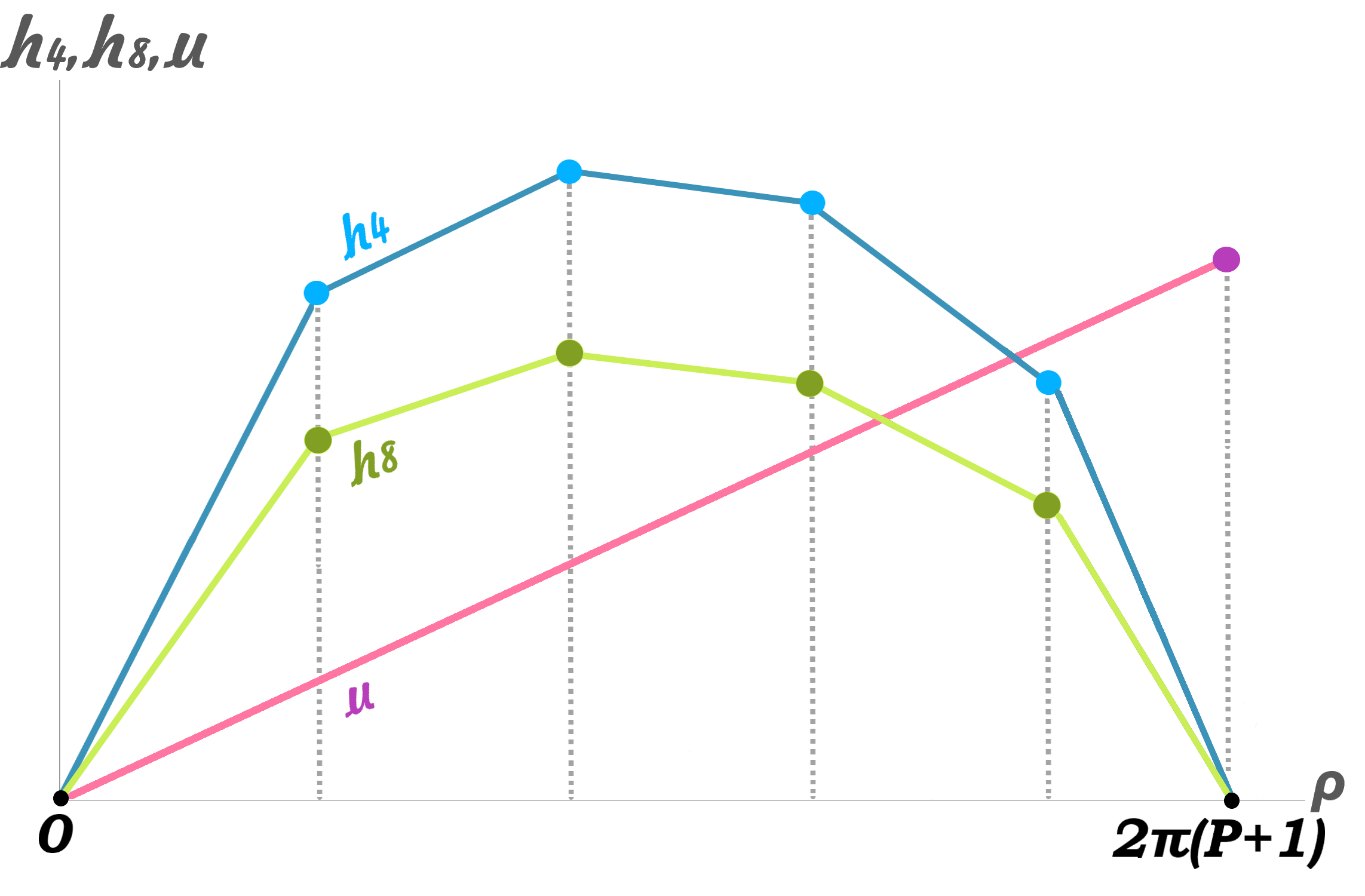} }}%
   % \qquad
    %\subfloat[label 2]
    %{{\includegraphics[width=4.5cm]{quiver_1_alpha_p.jpg} }}%
    %\qquad
    %\subfloat[label 3]
    %{{\includegraphics[width=4.5cm]{quiver_1_alpha_pp.jpg} }}
   % \caption{2 Figures side by side}%
\caption{An example of piecewise linear functions $h_4,h_8$ and of $u$, defining a particular supergravity background. Here, both $h_4$ and $h_8$ vanish at the endpoints of the $\rho$-dimension.}
\label{figure1}
\end{figure}

Nonetheless, the defining functions have to be chosen with some care for the space to properly close on the $\rho$-dimension. Considering a linear $u$ function, both $h_4,h_8$ need to be zero at the $\rho=0$ endpoint whereas at $\rho=2\pi(P+1)\equiv\rho_f$ only one of them needs to vanish. For a constant $u$ function, on the other hand, just one of them has to vanish at any endpoint. The study in \cite{Lozano:2019jza,Lozano:2019zvg} focused exclusively on solutions where both of these defining functions vanish at the endpoints, i.e. for $\alpha_0=\mu_0=a=0$ and $\nu_P=-\mu_P, \beta_P=-\alpha_P$ in the above definitions (\ref{h4piecewise}) and (\ref{h8piecewise}), a particular choice being represented by Figure \ref{figure1}. In Section \ref{Section3} of the present work, we investigate all other possible cases, where $h_4$ and $h_8$ generically do not vanish at the endpoints of the $\rho$-coordinate.

This particular choice of backgrounds $-$ where $h_4$ and $h_8$ are both zero at the endpoints of the $\rho$-dimension $-$ start in a smooth fashion on this coordinate as the non-Abelian T-duals of AdS$_3$ $\times$ S$^3$ $\times$ CY$_2$ \cite{Lozano:2019emq}. Near the endpoint $\rho=2\pi(P+1)-x$ with $x\rightarrow0$, on the other hand, the space becomes

\begin{equation}
\dd s^2\sim\frac{s_1}{x}\dd s^2_{\mbox{\tiny AdS$_3$}}+s_3\dd s^2_{\mbox{\tiny CY$_2$}}+\frac{x}{s_1}\left(\dd x^2+s_1s_2\dd s^2_{\mbox{\tiny S$^2$}}\right)\:,\hspace{1cm}e^{-4\phi}=s_4x^2\,,
\end{equation}\\
where $s_i$ are constants. According to the extremal $p$-brane solutions, classified in Appendix \ref{appendixA}, this space is a superposition of O2/O6 planes, where the O2 are smeared over O6.

In order to gain a better grip on the parameters of the system, let us consider the RR charges on the intervals $[2\pi k,2\pi(k+1)]$. For $\alpha'=g_s=1$, a D$p$-brane is charged under $Q_{Dp}=(2\pi)^{p-7}\int_{\Sigma_{8-p}}\hat{F}_{8-p}$, thus in our set-up they read

\begin{equation}
\begin{split}
Q_{D2}\;&=\;\frac{1}{32\pi^5}\int_{\mbox{\scriptsize CY$_2\times$S$^2$}}\hat{F}_6\;=\;h_4-h_4'(\rho-2\pi k)\;=\;\alpha_k\,,\hspace{1cm}Q_{D4}\;=\;\frac{1}{8\pi^3}\int_{\mbox{\scriptsize CY$_2$}}\hat{F}_4\;=\;\beta_k\,,\\
Q_{D6}\;&=\;\frac{1}{2\pi}\int_{\mbox{\scriptsize S$^2$}}\hat{F}_2\;=\;h_8-h_8'(\rho-2\pi k)\;=\;\mu_k\,,\hspace{2.1cm}Q_{D8}\;=\;2\pi F_0\;=\;2\pi h_8'\;=\;\nu_k\,,
\end{split}\label{PAGEcharges}
\end{equation}
Also, $Q_{NS}=\frac{1}{4\pi^2}\int_{\rho\times\mbox{\scriptsize S}^2}H_3=1$, while we used that vol(CY$_2)=16\pi^4$. These results imply that $\alpha_k,\beta_k,\mu_k,\nu_k$ are integers. A study of the Bianchi identities in the next section reveals that no explicit D2 and D6 branes are present in the geometry, just their fluxes\footnote{This is true when the worldvolume gauge field on the D8, D4 branes is absent. When it is on, as we are about to see, there is D6 and D2 flavor charge induced on the D8's and D4's.}. This associates their amount, $\alpha_k$ and $\mu_k$ respectively, with the ranks of the (color) gauge groups in the dual field theory. On the other hand, as restated, D8 and D4 branes do exist in the geometry and modify the Bianchi identities by a delta function. Thus, $\beta_k$ and $\nu_k$ are associated with the ranks of the (flavor) global symmetries of the dual field theory.\\

\subsection{Bianchi identities}
The above story is conjectured \cite{Lozano:2019jza,Lozano:2019zvg,Lozano:2019ywa} to be generated by a certain Hanany-Witten brane set-up \cite{Hanany:1996ie}. However, in this case the D-branes are not distributed across flat space as usual but along flat dimensions and a CY$_2$ manifold instead, as indicated by Table \ref{table1}.

\begin{table}[t!]
    \centering
    %\subfloat[label 1]
    {{\includegraphics[width=9cm]{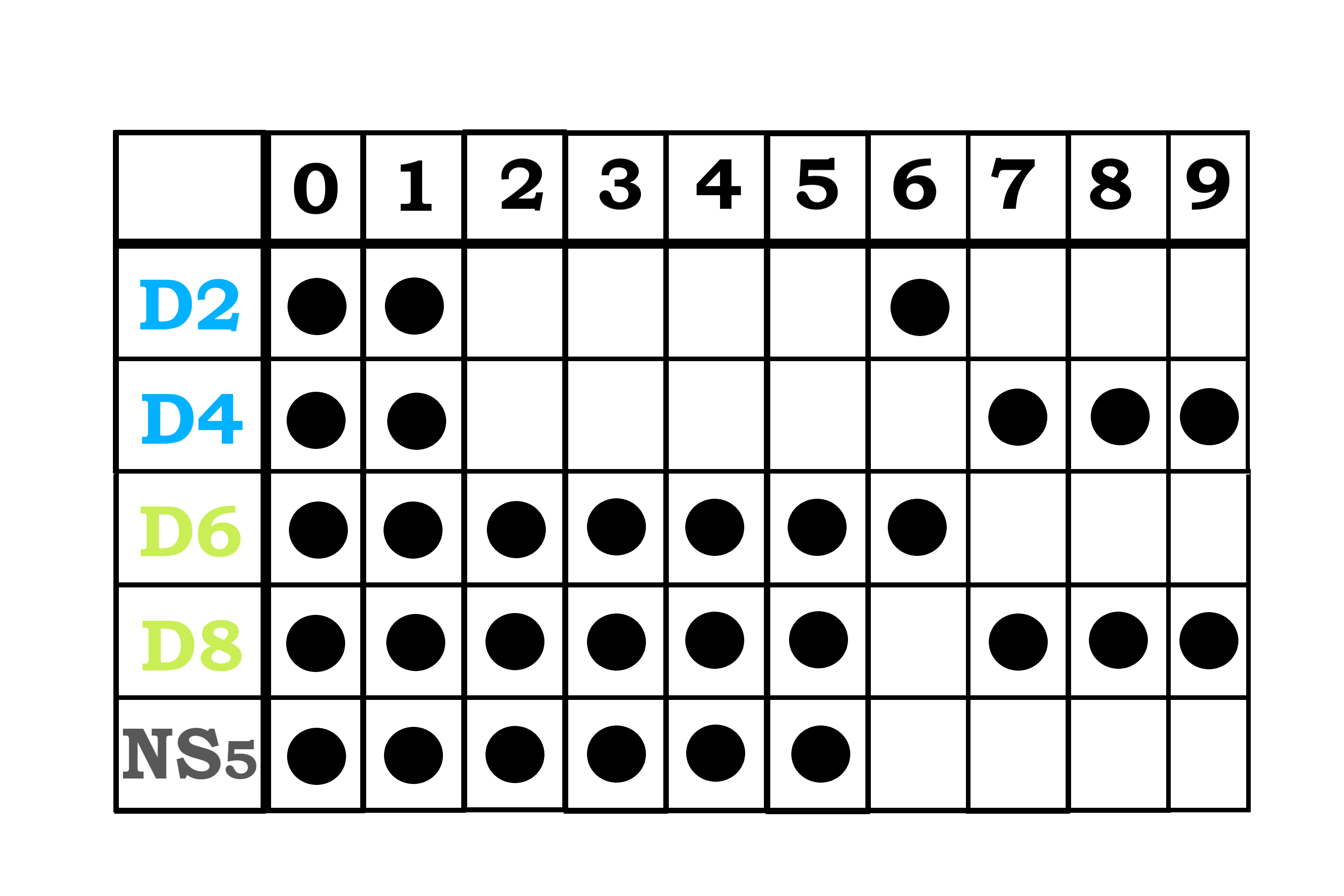} }}%
   % \qquad
    %\subfloat[label 2]
    %{{\includegraphics[width=4.5cm]{quiver_1_alpha_p.jpg} }}%
    %\qquad
    %\subfloat[label 3]
    %{{\includegraphics[width=4.5cm]{quiver_1_alpha_pp.jpg} }}
   % \caption{2 Figures side by side}%
\caption{$\frac{1}{8}$-BPS brane set-up, generator of our supergravity backgrounds. The dimensions $(x_0,x_1)$ are where the 2d CFT lives. The dimensions $(x_2,...,x_5)$ span the CY$_2$, on which the D6 and the D8-branes are wrapped. The coordinate $x_6$ is associated with $\rho$. Finally $(x_7,x_8,x_9)$ are the transverse directions realizing an SO(3)-symmetry associated with the isometries of S$^2$.}
\label{table1}
\end{table}

The family of supergravity backgrounds (\ref{GeneralAdS3background}) comes to be as the near-horizon limit of this brane set-up. Nevertheless, not all D-branes are explicitly present in the near-horizon limit of a Hanany-Witten set-up; some are there while others exist only as RR fluxes. This distinction is immensely important to Section \ref{Section3} and, thus, to clarify the situation we turn our attention to the Bianchi identities.

We begin by noticing that $\dd F_0=h_8''\dd\rho$ and $\dd \hat{F}_4=h_4''\dd\rho\wedge\mbox{vol(CY$_2$)}$ where, as reflected on the equations (\ref{SUGRAeom}), $h_4''=h_8''=0$ at a generic point along $\rho$. However, $h_4$ and $h_8$ are piecewise functions, given by (\ref{h4piecewise}) and (\ref{h8piecewise}), which means that at the points where their slope changes we get

\begin{equation}
h_4''=\sum_{k=1}^P\left(\frac{\beta_{k-1}-\beta_k}{2\pi}\right)\delta(\rho-2k\pi)\,,\hspace{1cm}h_8''=\sum_{k=1}^P\left(\frac{\nu_{k-1}-\nu_k}{2\pi}\right)\delta(\rho-2k\pi)\,.\label{h''}
\end{equation}
These give the source equations

\begin{equation}
\begin{split}
\dd F_0\;&=\;h_8''\,\dd\rho\,,\hspace{3cm}\dd\hat{F}_6\;=\;\dd\hat{f}_6\;=\;\frac{1}{2}h_4''\,(\rho-2k\pi)\,\dd\rho\wedge\mbox{vol(S$^2$)}\wedge\mbox{vol(CY$_2$)}\,,\\[10pt]
\dd \hat{F}_4\;&=\;\dd \hat{f}_4\;=\;h_4''\,\dd\rho\wedge\mbox{vol(CY$_2$)}\,,\hspace{1cm}\dd\hat{F}_2\;=\;\dd\hat{f}_2\;=\;\frac{1}{2}h_8''\,(\rho-2k\pi)\,\dd\rho\wedge\mbox{vol(S$^2$)}\,,
\end{split}\label{BianchisNormal}
\end{equation}
indicating that there are localized D4 and/or D8 branes at points $\rho=2k\pi$, whenever the slope between the intervals $[k-1,k]$ changes. In fact, the D4-branes are smeared over CY$_2$, while note that $f_p$ represents the magnetic part of a RR flux $F_p$. We also used that $x\delta(x)=0$, which yields that there are no sources present for the D6 and D2-branes. This is because of the large gauge transformations of the Kalb-Ramond field.

The above source equations suggest that the D2 and D6-branes play the role of color branes, while the D4 and D8-branes that of flavor branes. Since gauge transformations vanish at infinity, it is the gauge fields fluctuating on the D4 or D8-branes in the bulk that are realized as global (flavor) symmetries in the dual field theory. Ultimately, the essential feature of the Bianchi identities which becomes crucial in the forthcoming analysis is that the derivatives of $h_4$ and $h_8$ source D4 and D8-branes, respectively.

In the above source equations, however, we have not considered the gauge fields living on the D4 and D8 branes. Switching on a gauge field $\tilde{f}_2$ on both kinds of D-branes, we form the gauge invariant field strength $\mathcal{F}_2=B_2+\lambda\tilde{f}_2$, where $\lambda=2\pi l_s^2$, and the Bianchi identities now become
\begin{equation}
\begin{split}
\dd \hat{f}_2\;&=\;\lambda\tilde{f}_2\wedge\dd F_0\,,\\[10pt]
\dd \hat{f}_4\;&=\;h_4''\dd\rho\wedge\mbox{vol(CY$_2$)}\,+\,\frac{\lambda^2}{2}\tilde{f}_2\wedge\tilde{f}_2\wedge\dd F_0\,,\\[10pt]
\dd \hat{f}_6\;&=\;\lambda\tilde{f}_2\wedge\left(h_4''\dd\rho\wedge\mbox{vol(CY$_2$)}\right)\,+\,\frac{\lambda^3}{3!}\tilde{f}_2\wedge\tilde{f}_2\wedge\tilde{f}_2\wedge\dd F_0\,.
\end{split}\label{Bianchis+GFields}
\end{equation}\\
In regard to the gauge field dynamics, it being of order $l_s^2$, one may neglect it and keep only the zeroth order contribution, that is the Bianchi identities (\ref{BianchisNormal}) that give only D8 and D4-branes; this is what was assumed in \cite{Lozano:2019jza}. In Section \ref{Section3} of the present work, however, we deal with cases where the gauge field does become important and completely redefines the supergravity picture on the boundaries of the space.\\

\subsection{$\mathcal{N}=(0,4)$ SCFT}\label{subsectionQUIVERtheory}
The conjecture of \cite{Lozano:2019zvg} is that the above family of supergravity backgrounds is dual to a set of two dimensional SCFTs with $\mathcal{N}=(0,4)$ supersymmetry. These SCFTs are considered to be the low energy fixed points on the RG flows of well defined quantum field theories. Here, we just introduce the basic idea on those better-understood UV particle theories, ultimately aiming to cancel gauge anomalies that shall arise and also to unravel some interesting properties of the quiver superpotential.

\subsubsection{Gauge and global anomalies}
The quiver gauge theory of \cite{Lozano:2019zvg} may be outlined by its fundamental building block of superfields, given by Figure \ref{fig_block}. The field content and action of those multiplets is given in Appendix \ref{appendixFields} and, besides giving basic insight on the quiver structure, it is used in Section \ref{Section5} to build an operator and challenge its interacting properties.

\begin{figure}[t!]
    \centering
    %\subfloat[label 1]
    {{\includegraphics[width=9cm]{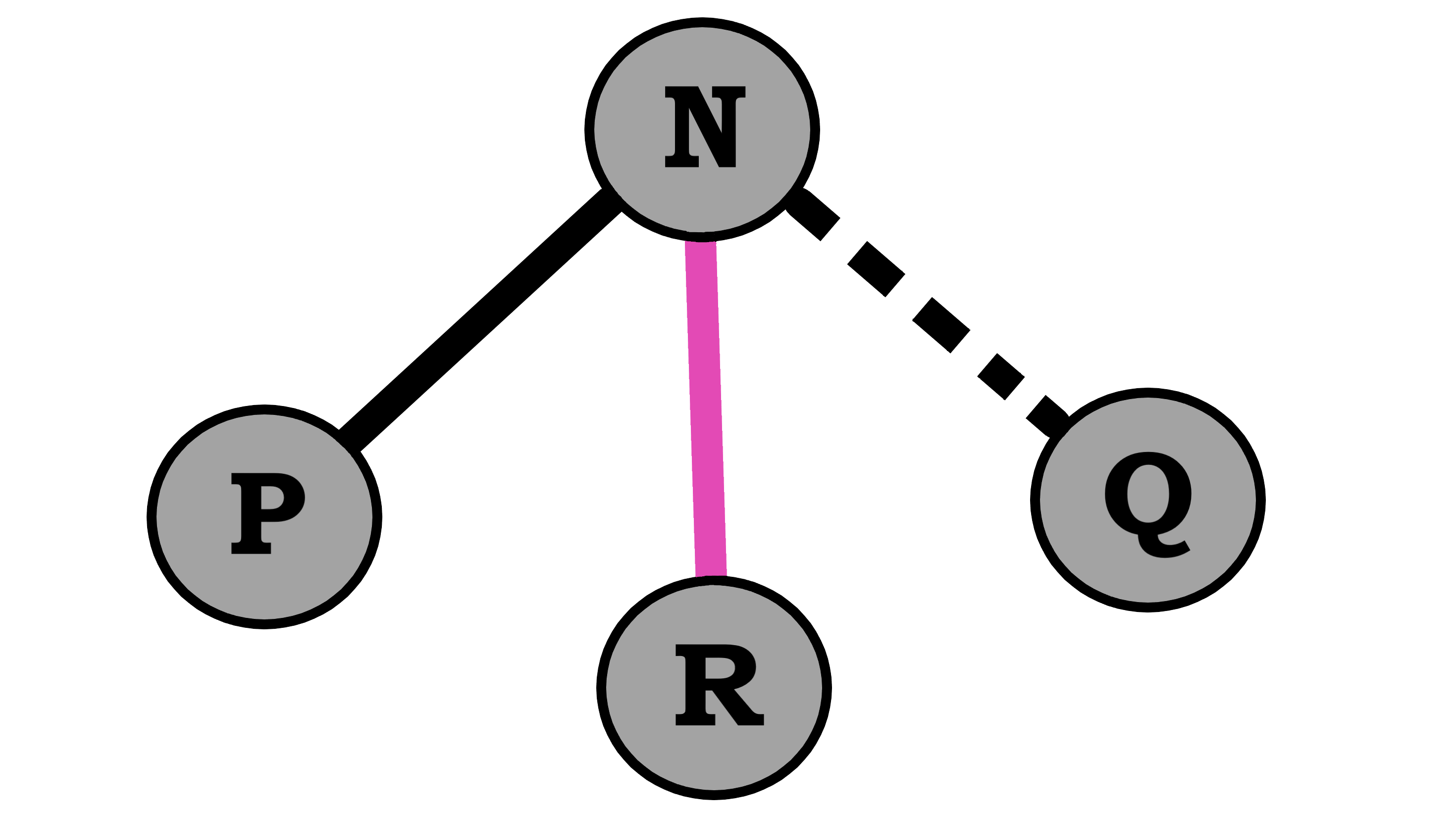} }}%
   % \qquad
    %\subfloat[label 2]
    %{{\includegraphics[width=4.5cm]{quiver_1_alpha_p.jpg} }}%
    %\qquad
    %\subfloat[label 3]
    %{{\includegraphics[width=4.5cm]{quiver_1_alpha_pp.jpg} }}
   % \caption{2 Figures side by side}%
\caption{The building block of our quiver field theories. The solid black line represents a $\mathcal{N}=(4,4)$ hypermultiplet, the maroon line a $\mathcal{N}=(0,4)$ hypermultiplet and the dashed line represents a $\mathcal{N}=(0,2)$ Fermi multiplet. Inside the node representing an SU$(N)$ gauge theory lives a $\mathcal{N}=(4,4)$ vector multiplet. The groups SU$(P)$, SU$(Q)$ and SU$(R)$ can be gauge or global symmetries.}
\label{fig_block}
\end{figure}

Each SU$(N)$ gauge theory living on $N$ D2 or D6 color branes is represented by a gauge node that yields a $\mathcal{N}=(4,4)$ vector multiplet. In $\mathcal{N}=(0,2)$ language, each gauge node includes a vector, a Fermi and two twisted chiral multiplets in the adjoint representation of SU$(N)$. A gauge node connects with other (gauge or flavor) nodes which in turn represent theories of (gauge or global) symmetry groups SU$(P)$, SU$(R)$ and SU$(Q)$, providing altogether a quiver network that reflects superstrings suspended between branes.

In the notation of Figure \ref{fig_block}, the SU$(N)$ gauge node connects to the SU$(P)$ (gauge or flavor) node through a $\mathcal{N}=(4,4)$ hypermultiplet. In $\mathcal{N}=(0,2)$ language, each such hypermultiplet includes two Fermi and two chiral multiplets. Since there are $NP$ kinds of strings between the SU$(N)$ and the SU$(P)$ brane stacks, we realize $2NP$ of each of these Fermi and chiral multiplets. The SU$(N)$ gauge node also connects to a SU$(R)$ node, through a $\mathcal{N}=(0,4)$ hypermultiplet. That is, through two $\mathcal{N}=(0,2)$ chiral multiplets. Since there are $NR$ kinds of strings between the SU$(N)$ and the SU$(R)$ brane stacks, we realize $2NR$ chiral multiplets connecting the two nodes. In the same manner, the SU$(N)$ gauge node connects to a SU$(Q)$ node, through $NQ$ $\mathcal{N}=(0,2)$ Fermi multiplets. 

All that being said, we may consider the superfield content of Appendix \ref{appendixFields} to find the overall anomaly of the gauge group SU$(N)$ and impose that it cancels, the result given by
\begin{equation}
2R\;=\;Q\label{anomalyCond}
\end{equation}
which analogously must hold for each gauge group in a consistent quiver gauge theory.

Non critical for the consistency of the gauge theory but as much essential to our analysis is the anomaly produced by the R-symmetry current. Focusing on the SU$(N)$ gauge theory of our building block and considering the U$(1)_R$ R-charges that are given in Appendix \ref{appendixRanomaly}, we find that the total R-anomaly reads $\Tr[\gamma_3Q_i^2]\sim 2(n_{hyp}-n_{vec})$ which is proportional to the difference between the hypermultiplets and the vector superfields of the building block. As derived in \cite{Lozano:2019jza,Putrov:2015jpa} this anomaly is linked to the central charge of the theory
\begin{equation}
c\;=\;6\left(n_{hyp}-n_{vec}\right)
\end{equation}
which will be vital in Section \ref{Section4}, where we want to add matter in the theory while leaving this charge intact.

\subsubsection{Quiver superpotential}\label{subsectionSuperpotential}
As promised, we now realize a superpotential on our quiver theory by focusing on its building block given by Figure \ref{fig_block}. In particular, we just take one simple connection of it, that is the link between a hypermultiplet and a vector superfield. All other links on the quiver structure can be deduced as generalizations of this connection. In fact, a particular two-dimensional superpotential was developed in \cite{Tong:2014yna} that serves exactly our case; we briefly reproduce this here, in order to extract the field interactions which furnish a certain operator in Section \ref{Section5} with special features.

Through $\mathcal{N}=(0,2)$ supersymmetric eyes, a $\mathcal{N}=(4,4)$ vector superfield breaks into a vector multiplet $\mathcal{V}$, a Fermi multiplet $\Theta$ and two (twisted) chiral multiplets $\Sigma,\tilde{\Sigma}$. On the other hand, a $\mathcal{N}=(4,4)$ hypermultiplet breaks into two chiral multiplets $\Phi,\tilde{\Phi}$ and two Fermi multiplets $\Gamma,\tilde{\Gamma}$. First things first, considering transformation properties under the R-symmetry, the Fermi multiplet $\Theta$ inside the vector superfield may only be defined through $\bar{\mathcal{D}}_+\Theta\;=\;E_\Theta$ by the holomorphic function

\begin{equation}
E_\Theta\;=\;[\Sigma,\tilde{\Sigma}]
\end{equation}\\
and by the superpotential $\mathcal{W}_\Theta=\tilde{\Phi}\Theta\Phi$, where $J_\Theta=\tilde{\Phi}\Phi$ is another holomorphic function.

On the contrary, the R-symmetry representations furnishing the $\mathcal{N}=(4,4)$ hypermultiplet, define its Fermi multiplets as
\begin{equation}
E_\Gamma\;=\;\Sigma\Phi\,,\hspace{2cm}E_{\tilde{\Gamma}}\;=\;-\tilde{\Phi}\Sigma
\end{equation}
and let for the superpotential $\mathcal{W}_\Gamma+\mathcal{W}_{\tilde{\Gamma}}=\tilde{\Phi}\tilde{\Sigma}\Gamma+\tilde{\Gamma}\tilde{\Sigma}\Phi$, where $J_\Gamma=\tilde{\Phi}\tilde{\Sigma}$ and $J_{\tilde{\Gamma}}=\tilde{\Sigma}\Phi$.

In reality, it is not just the R-symmetry representations that we took into account to shape the above functions, but also the constraining condition $E\cdot J=\sum_aE_aJ^a=0$ that should hold for supersymmetry to be preserved; of course, it is easy to see that this is satisfied for the given functions. The holomorphic functions $E_a$ and $J^a$ give the potential terms $\sim\abs{E_a(\phi_i)}^2$ and $\sim\abs{J_a(\phi_i)}^2$ in the action and produce an interesting interactive sector in our theory that is going to become decisively important in Section \ref{Section5}.\\

\section{Dielectric branes on the boundary}\label{Section3}
The case studied in \cite{Lozano:2019jza,Lozano:2019zvg} and in the previous section is dedicated to supergravity solutions defined by functions $h_4,h_8$ that vanish at the endpoints of the $\rho$-dimension, as in Figure \ref{figure1}. Nevertheless, this is just one choice among many.\\

To classify all other possible kinds of solutions we must first consider the restrictions that apply on the functions $h_4,h_8$ and $u$. That is, these defining functions have to be chosen in such a way that the space properly closes on the $\rho$-dimension. Considering a linear $u$ function, both $h_4,h_8$ need to be zero at the $\rho=0$ endpoint whereas at $\rho=\rho_f$ only one of them needs to vanish. For a constant $u$ function, on the other hand, just one of them has to vanish at any endpoint. As we are about to find out, the physical set-up significantly changes depending on whether the function $u$ is linear or just a constant, both being legitimate solutions of the BPS equation $u''(\rho)=0$.

While all those \textit{novel} cases are totally valid as supergravity solutions (i.e. they satisfy the equations of motion (\ref{SUGRAeom})), a particular ambiguity arises in their dual quiver field theories. The ambiguity is that the gauge anomalies for these new quivers do not \textit{seem} to cancel. In particular, it is the color nodes on the edges of the quivers that $-$ naively $-$ seem anomalous.

A promising answer to this riddle arises by focusing back on the supergravity side and observing the limiting geometry at the endpoints of the $\rho$-dimension (where the physics is dual to the aforementioned color nodes at the quiver edges). On those limiting vicinities, in contrast with the original paradigm of the previous section where the limiting space is either smooth or has O-planes, we now find D-branes. This is promising because explicit D-branes correspond to flavor symmetries (i.e. flavor nodes) that may contribute in the necessary way to cancel the gauge anomalies. Indeed, this is exactly what happens. But let us better realize all this through some solid examples.\newpage

\begin{figure}
\centering
\begin{subfigure}[b]{0.48\textwidth}
 \centering
  \includegraphics[width=1\linewidth]{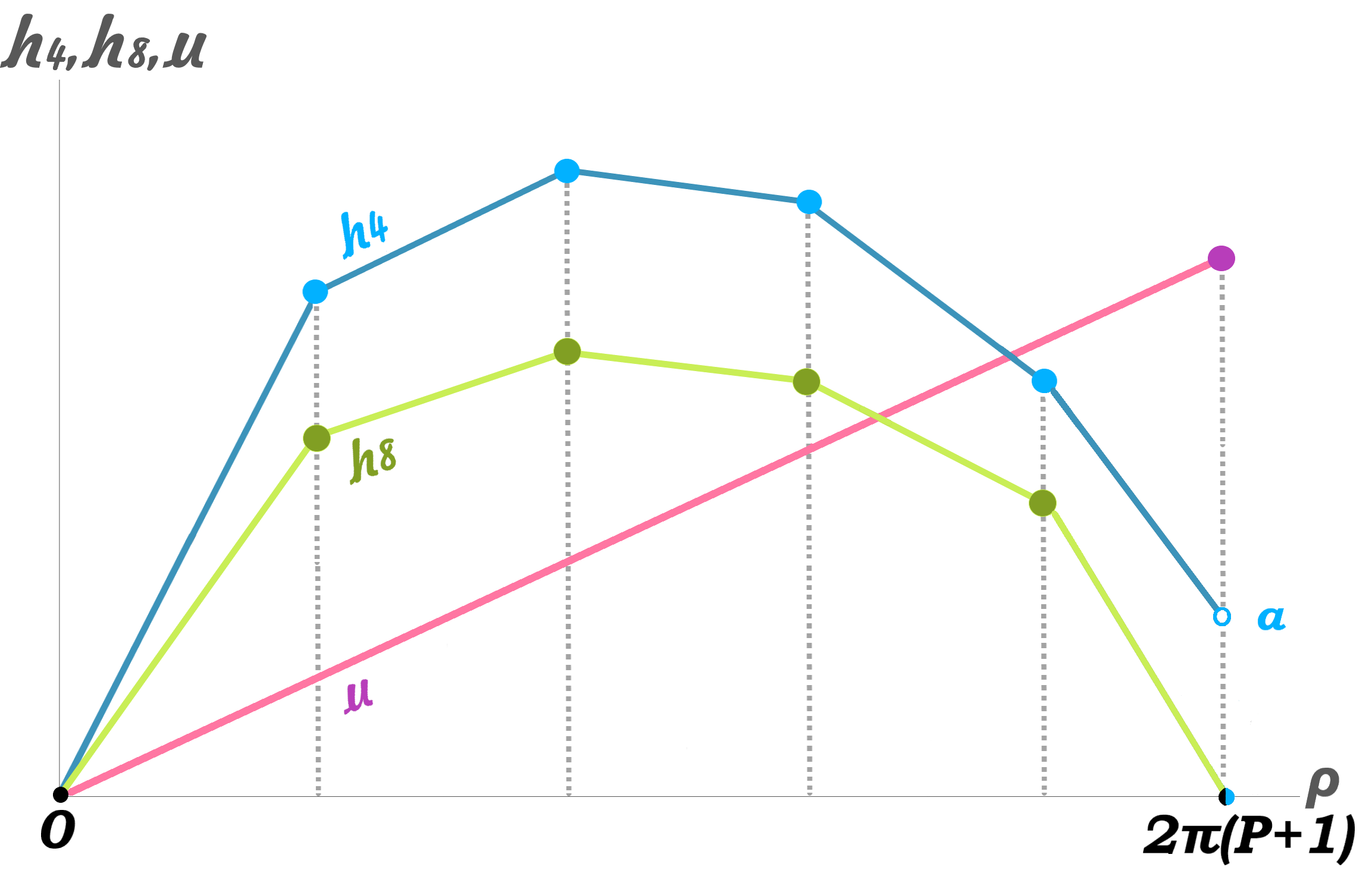}
  \caption{A background with linear $u$ and a non-vanishing $h_4$ at the endpoint.}
  \label{figure3a}
\end{subfigure}
\hspace{1em}%
\begin{subfigure}[b]{.48\textwidth}
  \centering
  \includegraphics[width=1\linewidth]{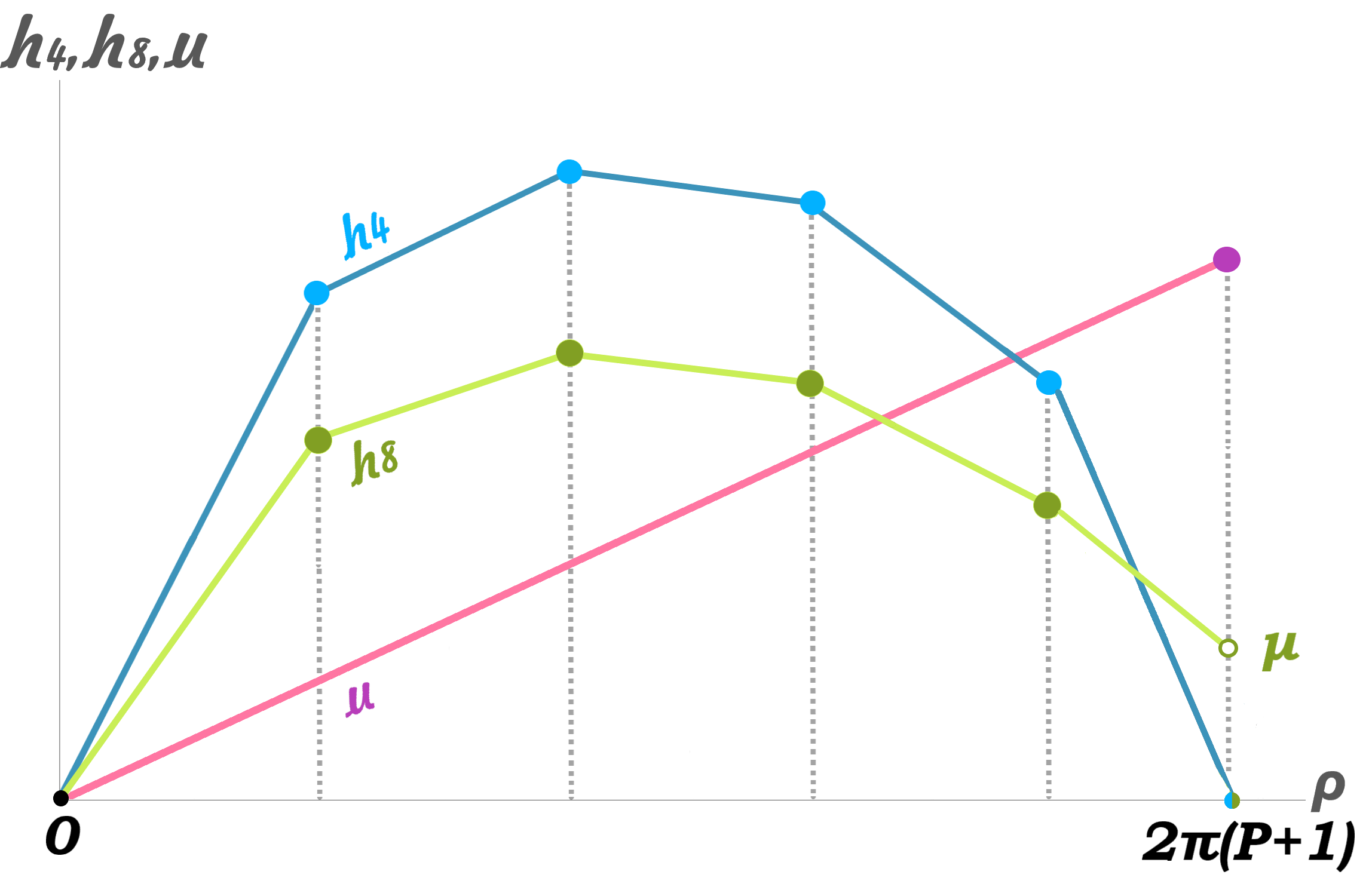}
  \caption{A background with linear $u$ and a non-vanishing $h_8$ at the endpoint.}
  \label{figure3b}
\end{subfigure}
\caption{All the possible classes of backgrounds defined by a linear function $u(\rho)$ and a non-vanishing function $h_4$ or $h_8$ at the endpoint $\rho=\rho_f$.}
\label{figurelinearu}
\end{figure}

\subsection{Linear $u(\rho)$}
As restated, the physics of the supergravity solutions changes depending on whether the function $u$ is linear or just a constant. Therefore, we split our analysis into two distinct parts, with regards to this property. The possible classes of backgrounds with linear $u$ and a non-vanishing $h_4$ or $h_8$ at the endpoint $\rho=\rho_f$ are classified in Figure \ref{figurelinearu}.\\

\subsubsection{Example I}
\begin{figure}
\centering
\begin{subfigure}[b]{0.48\textwidth}
 \centering
  \includegraphics[width=1\linewidth]{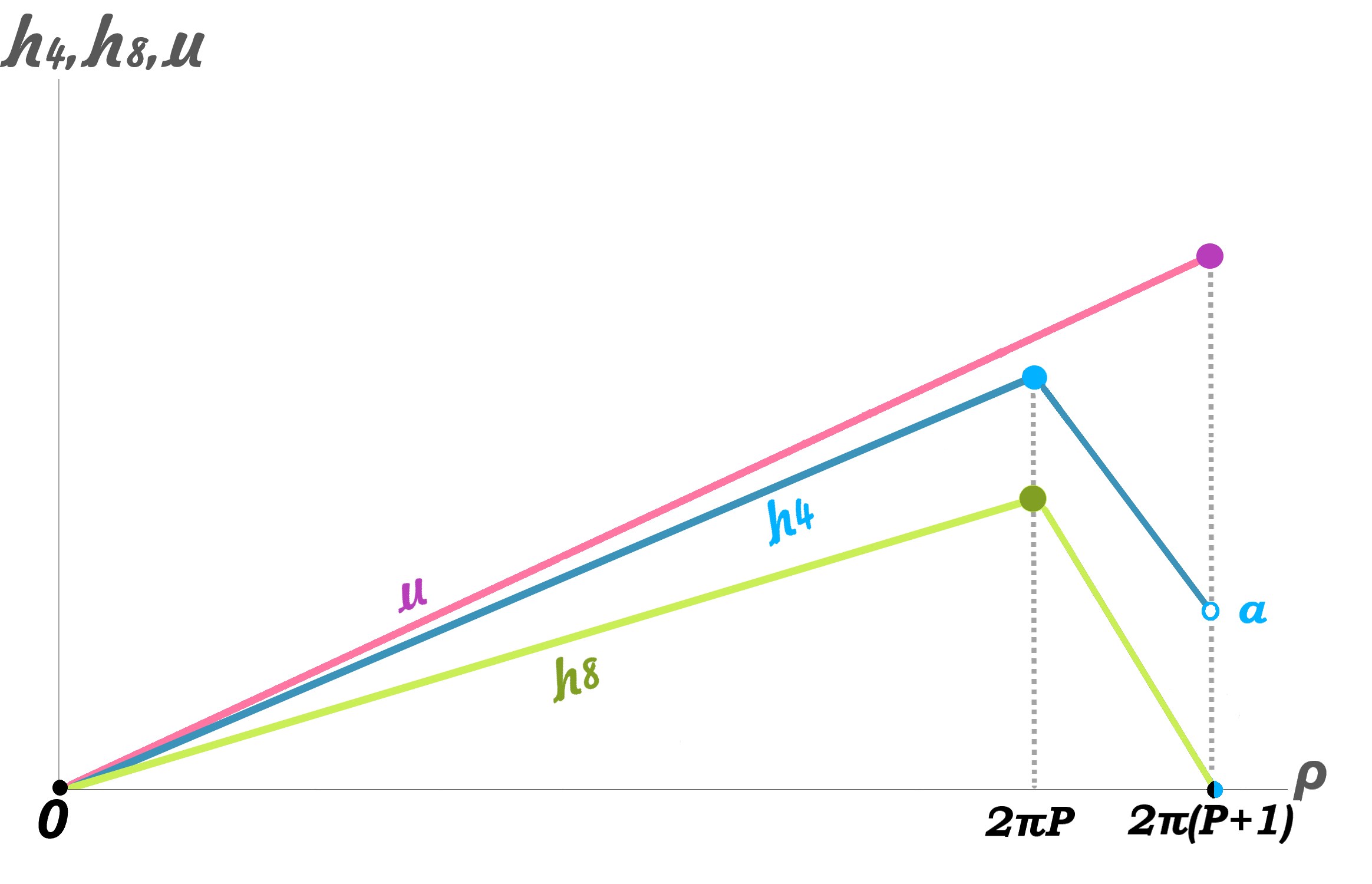}
  \caption{A simplified version of Figure \ref{figure3a}. The function $u$ is linear, $h_8$ starts and closes with a vanishing value and $h_4$ vanishes at zero but not at $\rho=\rho_f$.}
  \label{figure4a}
\end{subfigure}
\hspace{1em}%
\begin{subfigure}[b]{.48\textwidth}
  \centering
  \includegraphics[width=1\linewidth]{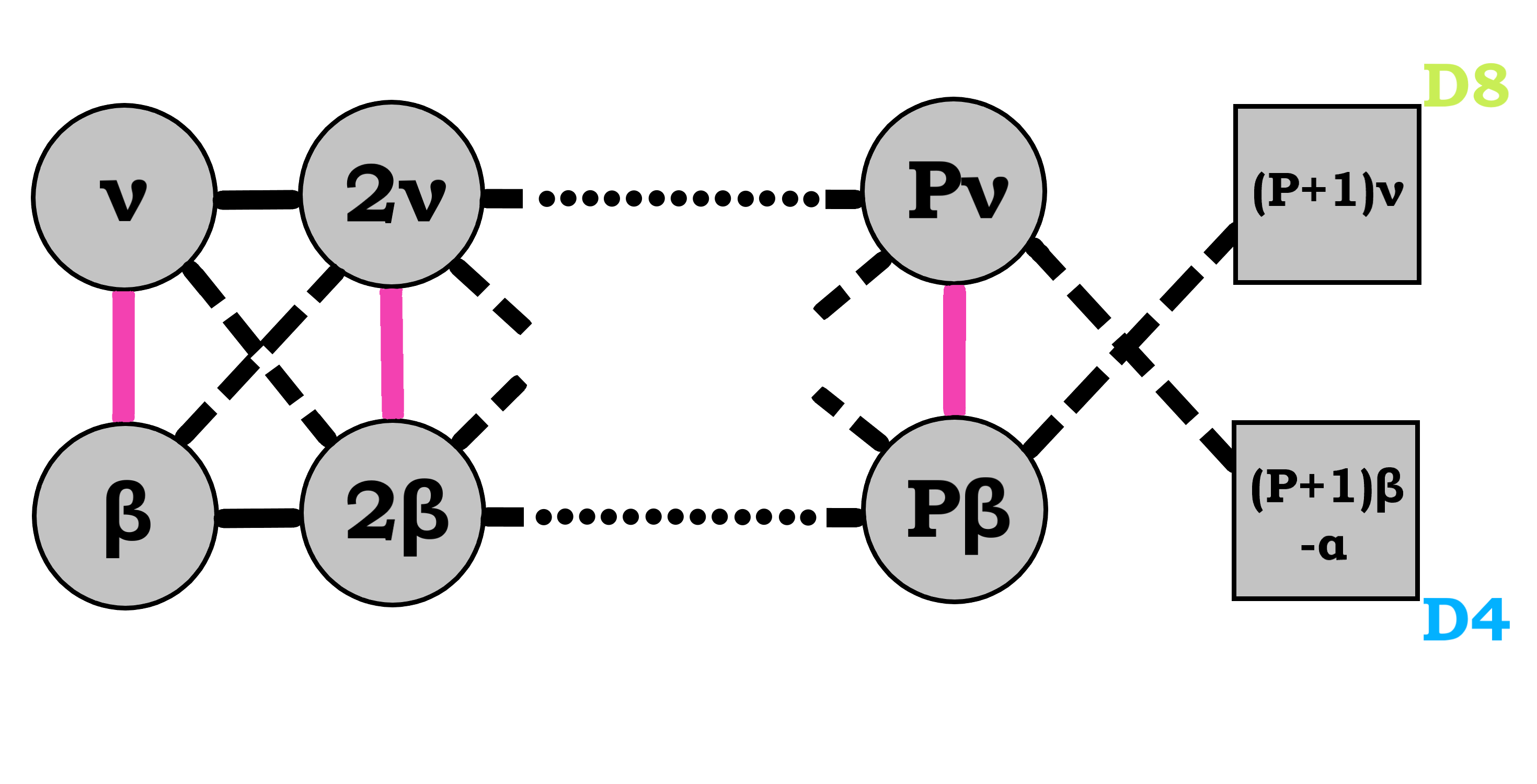}
  \caption{This is the naive quiver dual to the background defined by (\ref{h4example1}), (\ref{h8example1}). In reality, there is one more flavor node, canceling the gauge anomalies for the last D6 gauge node.}
  \label{figure4b}
\end{subfigure}
\caption{A simplified version of the background given in Figure \ref{figure3a} and its dual quiver theory. Here, besides a linear function $u$, $h_8$ starts and closes with a vanishing value, while $h_4$ starts at zero but finishes at a non-zero value.}
\label{figure4}
\end{figure}

We begin by studying the class of backgrounds that is defined by a linear function $u$ and a non-vanishing function $h_4$ at the endpoint $\rho=\rho_f$, that is Figure \ref{figure3a}. Nevertheless, because all the interesting action takes place in the last interval of the $\rho$-dimension (and its dual quiver gauge end-node) whose behavior we essentially care about, we shall study the simplest version of this class. That would be Figure \ref{figure4a}.

The class of backgrounds represented by Figure \ref{figure4a} are defined by a linear function $u$ and by the functions

\begin{equation}
h_4(\rho)\:=\:\left\lbrace\begin{array}{ccc}
\frac{\beta}{2\pi}\rho &\hspace{1cm}2\pi k\leq\rho\leq2\pi(k+1) &k=0,...,P-1\,,\\
\alpha-\frac{\beta P-\alpha}{2\pi}(\rho-2\pi(P+1)) &\hspace{1cm}2\pi P\leq\rho\leq2\pi(P+1) &
\end{array}\right.\label{h4example1}
\end{equation}
\begin{equation}
h_8(\rho)\:=\:\left\lbrace\begin{array}{ccc}
\frac{\nu}{2\pi}\rho &\hspace{2cm}2\pi k\leq\rho\leq2\pi(k+1) &k=0,...,P-1\,.\\
\frac{\nu P}{2\pi}(2\pi(P+1)-\rho) &\hspace{2cm}2\pi P\leq\rho\leq2\pi(P+1) &
\end{array}\right.\label{h8example1}
\end{equation}\\
The background defined by these functions is $-$ naively $-$ dual to the quiver theory given by Figure \ref{figure4b}. The fact that this quiver is not the right one can be easily seen by observing the last D6 gauge node, i.e. the one with gauge rank $P\nu$; using the anomaly cancellation condition (\ref{anomalyCond}), the gauge anomalies on this node do not cancel. On the contrary, anomaly cancellation would occur if the gauge node was to connect with an additional flavor node of rank $\alpha$ through a $\mathcal{N}=(0,2)$ Fermi multiplet.

This raises a puzzle, since the standard Hanany-Witten brane set-up introduced in \cite{Lozano:2019jza,Lozano:2019zvg} (and represented by Figure \ref{figure1}) does not include any additional D-branes at the endpoints of the $\rho$-dimension, which would support such an additional flavor symmetry. Nonetheless, in contrast to that particular case, our solution defined by (\ref{h4example1}) and (\ref{h8example1}) has the novelty of a non-vanishing function $h_4$ at $\rho=\rho_f$. Hence, we shall focus on that vicinity of the supergravity background, which is dual to the problematic D6 gauge node, and see whether there is anything interesting there. That is, we focus near the end point $\rho=2\pi(P+1)-x$, for $x\rightarrow0$, where the geometry and the dilaton read

\begin{equation}
\dd s^2\;=\; \frac{1}{\sqrt{x}}\left(s_1\,\dd s^2_{\mbox{\tiny AdS$_3$}}+s_2\,\dd s^2_{\mbox{\tiny CY$_2$}}\right)+\sqrt{x}\left(s_3\,\dd x^2+s_4\,\dd s^2_{\mbox{\tiny S$^2$}}\right)\:,\hspace{1cm}e^{\phi}=s_5\,x^{-\frac{3}{4}}\,,\label{exampleID6}
\end{equation}\\
with $s_i$ real constants. As foreseen, we reached an interesting outcome since this background corresponds to D6-branes on AdS$_3\times$CY$_2$ and smeared over S$^2$. To be exact, the above metric and dilaton also correspond to O6-planes, however only D6-branes can host open strings on their worldvolume and, thus, we only consider those to deduce global symmetries. That is, being explicit branes, these D6's contribute to the flavor structure of the quiver theory and, in principle, they should cancel the gauge anomalies on the last D6 gauge node.

On the other hand, the Bianchi identities yield no explicit D6-branes in our supergravity construction. According to the violation of these identities, the $h_4$ function $-$ that appears here to feed the boundary of the space with D6-branes $-$ may only give rise to D4-branes. Hence, since we do know we should have D4-branes at the endpoint where $h_4$ does not vanish, while we do not see them, we go on and study their sources. That is, we look upon their full Chern-Simons action \cite{Myers:1999ps}

\begin{equation}
\begin{split}
S_{\mbox{\tiny CS}}^{\mbox{\tiny D4}}\;&=\;\mu_4\int\Tr\:\sum e^{i\lambda\imath_\Phi\imath_\Phi} C_{(n)}e^{\mathcal{F}_2}\\[10pt]
&=\;\mu_4\int\Tr\:C_5^{el}+C_3^{el}\wedge\mathcal{F}_2+i\lambda(\imath_\Phi\imath_\Phi)C_7^{el}-\lambda^2(\imath_\Phi\imath_\Phi)^2\left(C_9^{el}+C_7^{el}\wedge\mathcal{F}_2+\ldots\right)
\end{split}\label{D4WZ}
\end{equation}\\
where the sum keeps only five-form terms that may source D4-branes. $C^{el}$ is the electric part of a potential form, $\mathcal{F}_2=B_2+\lambda\tilde{f}_2$ is the gauge invariant field strength that incorporates the D4 worldvolume gauge field and $\imath_\Phi$ reflects the inner product with the D4-brane transverse modes $\Phi^i$. Dimensional analysis here implies $\lambda=2\pi l_s^2$. The first term in the second line sources standard D4-branes, the second term reflects a D4/D2 bound state, while the third gives a D4/D6 bound state and so on. While the object $C_{3}\wedge\mathcal{F}_2$ realizes D2-charge induced into the D4-brane worldvolume, the seminal work by Myers \cite{Myers:1999ps} showed that an RR potential coupled to the transverse modes $\Phi^i$ represents a polarization of lower-dimensional D-branes into a higher-dimensional one.

Taking into account the RR fluxes of (\ref{RRsector}) and the functional forms (\ref{h4example1}),(\ref{h8example1}) near the endpoint $\rho\rightarrow\rho_f$, we pick a convenient gauge choice and deduce that

\begin{equation}
\begin{split}
C_3^{el},C_5^{el}\;\;&\rightarrow\;\;const.\,,\hspace{2cm}C_7^{el}\;\propto\;\left(\frac{-1}{\rho_f-\rho}\right)\,\mbox{vol(AdS$_3$)}\wedge\mbox{vol(CY$_2$)}\;\;\rightarrow\;\;-\infty\,,\\[10pt]
C_9^{el}\;&\propto\;\left(\log(\rho_f-\rho)\right)\,\mbox{vol(AdS$_3$)}\wedge\mbox{vol(CY$_2$)}\wedge\mbox{vol(S$^2$)}\;\;\rightarrow\;\;-\infty\;.
\end{split}
\end{equation}\\

Since $C_7^{el}$ and $C_9^{el}$ blow up at the boundary, then their corresponding source terms in the Chern-Simons action (\ref{D4WZ}) dominate the game as opposed to the rest. Between those two potentials, $C_7^{el}$ scales infinitely faster as we approach $\rho_f$ and therefore we argue that, at the boundary, the D4-branes couple to an infinitely strong $C_7^{el}$ RR potential and condense out into D6-branes, yielding the analogous background (\ref{exampleID6}). In fact, it should be the fifth term in the expansion of (\ref{D4WZ}) that prevails; it is this particular term that yields bound states of D6-branes that are smeared over S$^2$ (under the coupling to $\mathcal{F}_2$), which agrees with the background (\ref{exampleID6}). The third term in (\ref{D4WZ}) gives just ordinary (not smeared) bound states of D6-branes\footnote{A more elaborate proof of this is based in the string length ($\lambda$-) order of those $C_7^{el}$-terms and comes through the analogous case of the upcoming Section \ref{Section3.2}, which is thoroughly analyzed in Appendix \ref{AppendixD8/D4}. There, we will show that only terms of, at least, order $\mathcal{O}(\lambda^2)$ can provide non-trivial solutions for the D-brane bound states.}. Finally, notice the fact that we have a non-vanishing $C_5^{el}$; this is vital for the very existence of the constituent D4-branes on the D4/D6 bound state.

Recalling our original goal, we want to find the way this D4/D6 bound state contributes to the flavor symmetry of the theory. That is, the strings on the condensed D4-branes form a U($N_4)$ gauge theory under certain conditions, $N_4$ being the number of those branes given by the Bianchi identity

\begin{equation}
\dd\hat{f}_4\;=\;h_4''\,\dd\rho\wedge\mbox{vol(CY$_2$)}\label{exampleID4bianchi}
\end{equation}\\
The U($N_4)$ flavor gauge group is what we are after and anticipate of it canceling the gauge anomalies in the quiver theory.

To calculate (\ref{exampleID4bianchi}) at the boundary, we have to handle things delicately. This is because the number of four-branes is associated with $h_4'$ and a derivative is not well defined on the endpoint of a closed interval. Therefore, we shall demand that $h_4|_{\rho_f}=0$, so that the derivative becomes well defined \textit{near} the endpoint $\rho_f$\footnote{The essence of differentiation is to realize how a function changes. In our particular context, the measure of this change is associated with the number of branes at a point. Since the background is defined on a closed interval, it makes sense to realize the absence of branes out of it as a shift of the defining function to a vanishing value. Stated otherwise, we exchange emptiness for a zero.}. This is not a physical requirement of any sort; it is just a trick to calculate the D-branes at the end of the space. Thus we now have the derivative

\begin{equation}
h_4'\Big|_{\rho\rightarrow\rho_f}\;=\;\lim_{x\rightarrow0}\frac{h_4(\rho_f)-h_4(\rho_f-x)}{x}=\lim_{x\rightarrow0}\frac{-\alpha}{x}\label{h4dev}
\end{equation}\\
and, in order to calculate all the four-branes on the endpoint, the D4 Page charge in (\ref{PAGEcharges}) has to be integrated\footnote{The trick we applied on the $h_4$ function, forms a situation where the branes appear smeared near the endpoint, instead of being localized with a delta function as with the rest of the D4-brane stacks along the $\rho$-dimension. This is merely an artifact of our particular handling that is resolved just by adding up (integrating over) all the branes near that endpoint.} towards $\rho_f$ as
\begin{equation}
N_4\;=\;-\int_{\rho_f-x}^{\rho_f} h_4'\;=\;\alpha
\end{equation}\\
Bottom line, we found $\alpha$ D4-branes sitting on the endpoint of the $\rho$-interval and being in a D4/D6 bound state.

The polarization that takes place should raise the question whether the D4-branes are enough in number, throughout the bound state, to support massless string modes and thus a unitary gauge theory. In reality, though, we are not obligated to know the precise geometry of the polarized branes, just that they are enough in number to be close to one another so that the modes do not get massive. And fortunately we do know that the D4-branes are a lot, since $\alpha$ must be large in the supergravity limit by construction. Therefore U$(\alpha)$ should be the gauge group we have anticipated.

\begin{figure}[t!]
    \centering
    %\subfloat[label 1]
    {{\includegraphics[width=11cm]{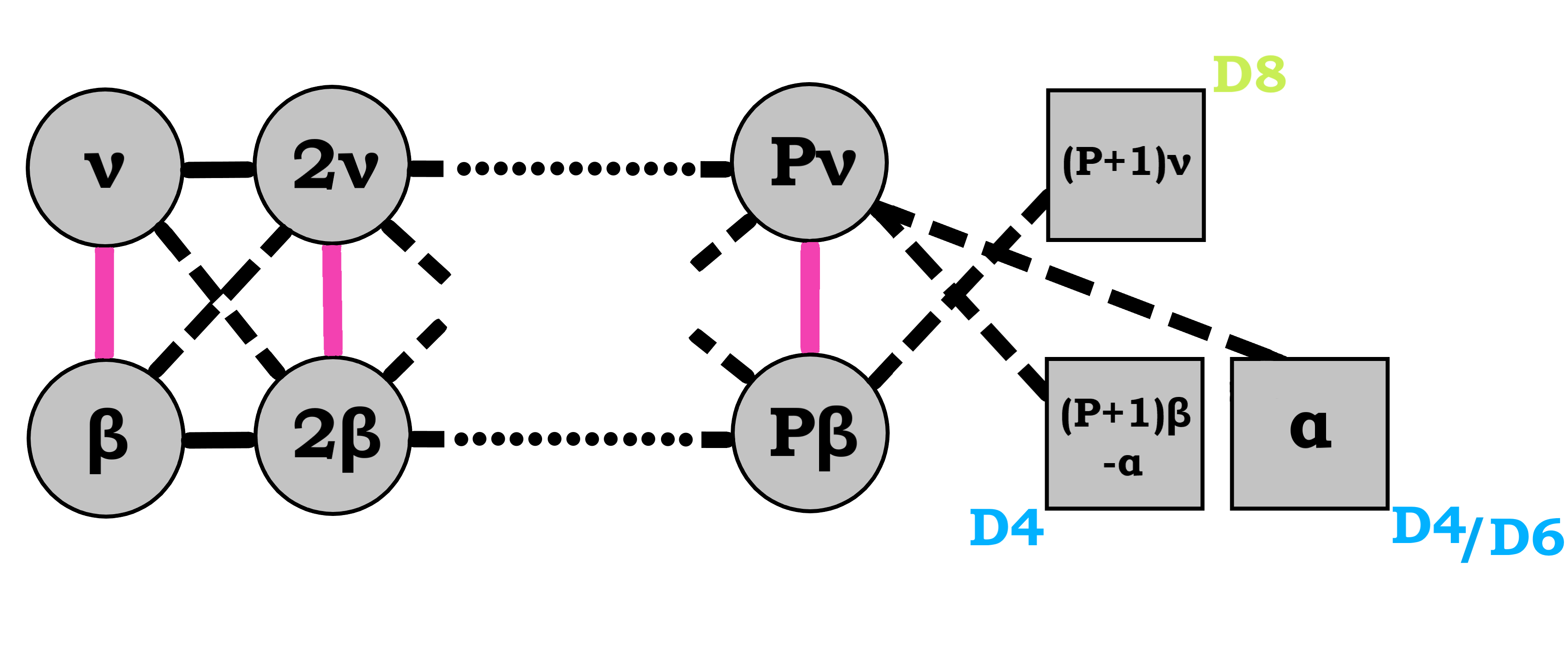} }}%
   % \qquad
    %\subfloat[label 2]
    %{{\includegraphics[width=4.5cm]{quiver_1_alpha_p.jpg} }}%
    %\qquad
    %\subfloat[label 3]
    %{{\includegraphics[width=4.5cm]{quiver_1_alpha_pp.jpg} }}
   % \caption{2 Figures side by side}%
\caption{This is the actual quiver dual to the background defined by (\ref{h4example1}), (\ref{h8example1}). Here, the extra four-brane flavor node cancels the gauge anomalies for the last $h_8$ (D6) gauge node.}
\label{figure10}
\end{figure}

Being explicit branes, the worldvolume theory of those D4-branes feeds, through a $\mathcal{N}=(0,2)$ Fermi multiplet, the D6 color chain of the quiver with flavor. In particular, this U($\alpha$) gauge group is dual to a global symmetry in the quiver theory which, using (\ref{anomalyCond}), gives exactly the flavor needed in order to cancel the gauge anomalies of the last D6 color chain node. This is all visualized in Figure \ref{figure10}, where the quiver theory is now consistent.\\

Focusing on the starting point $\rho=0$ of the $\rho$-interval, the background becomes the non-Abelian T-dual of AdS$_3\times$S$^3\times$CY$_2$, which yields no D-branes there. This is to be expected from the supergravity side, since everything is obviously smooth there. But even by just looking at the field theory, the quiver is non-anomalous at its beginning (and now everywhere for that matter), which means that no additional D-branes should be there. If there were any, these would contribute with flavor and spoil the anomaly cancellation balance.\\
\\

\subsubsection{Example II}
\begin{figure}
\centering
\begin{subfigure}[b]{0.48\textwidth}
 \centering
  \includegraphics[width=1\linewidth]{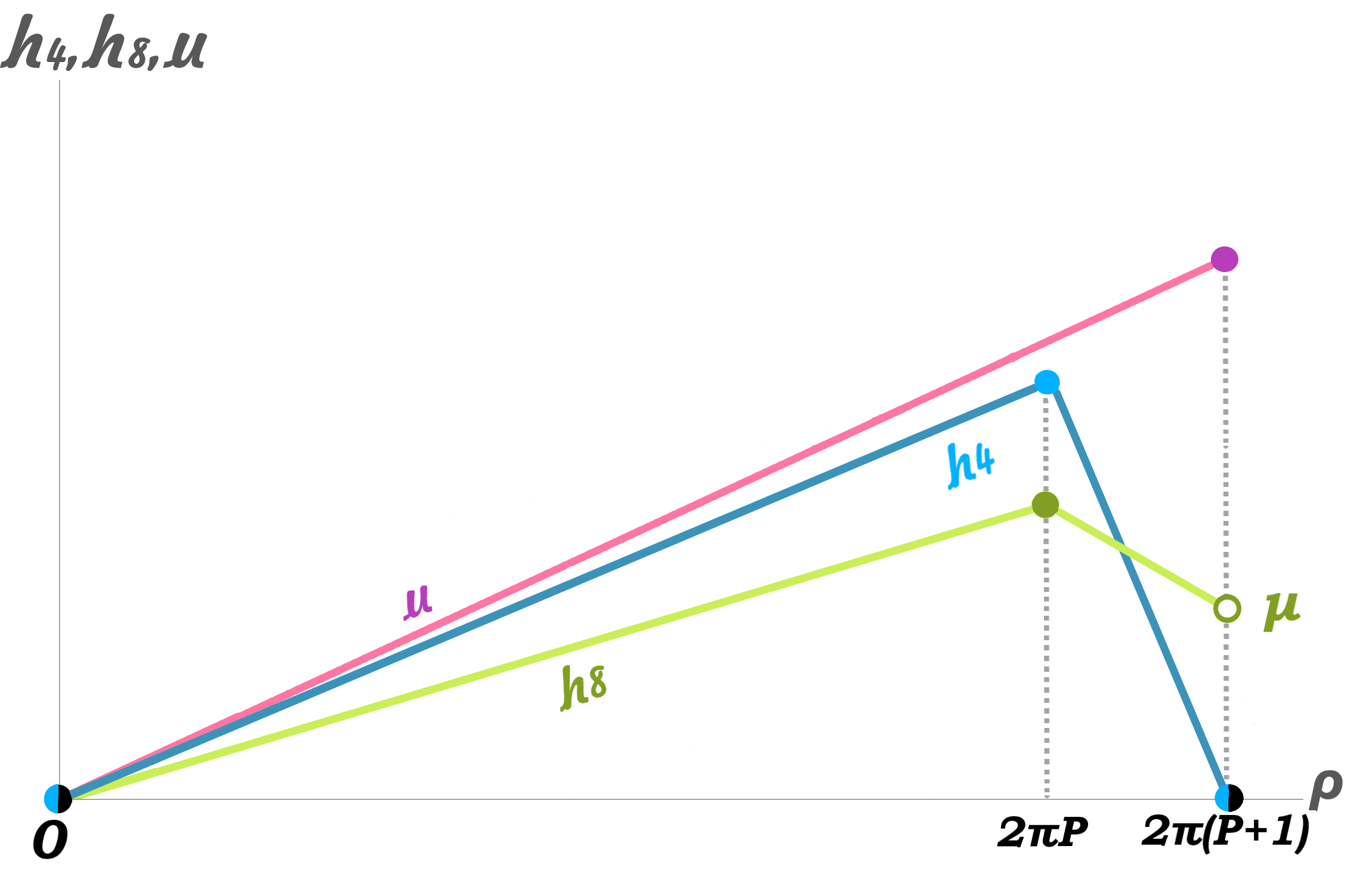}
  \caption{A simplified version of Figure \ref{figure3b}. The function $u$ is linear, $h_4$ starts and closes with a vanishing value and $h_8$ vanishes at zero but not at $\rho=\rho_f$.}
  \label{figure6a}
\end{subfigure}
\hspace{1em}%
\begin{subfigure}[b]{.48\textwidth}
  \centering
  \includegraphics[width=1\linewidth]{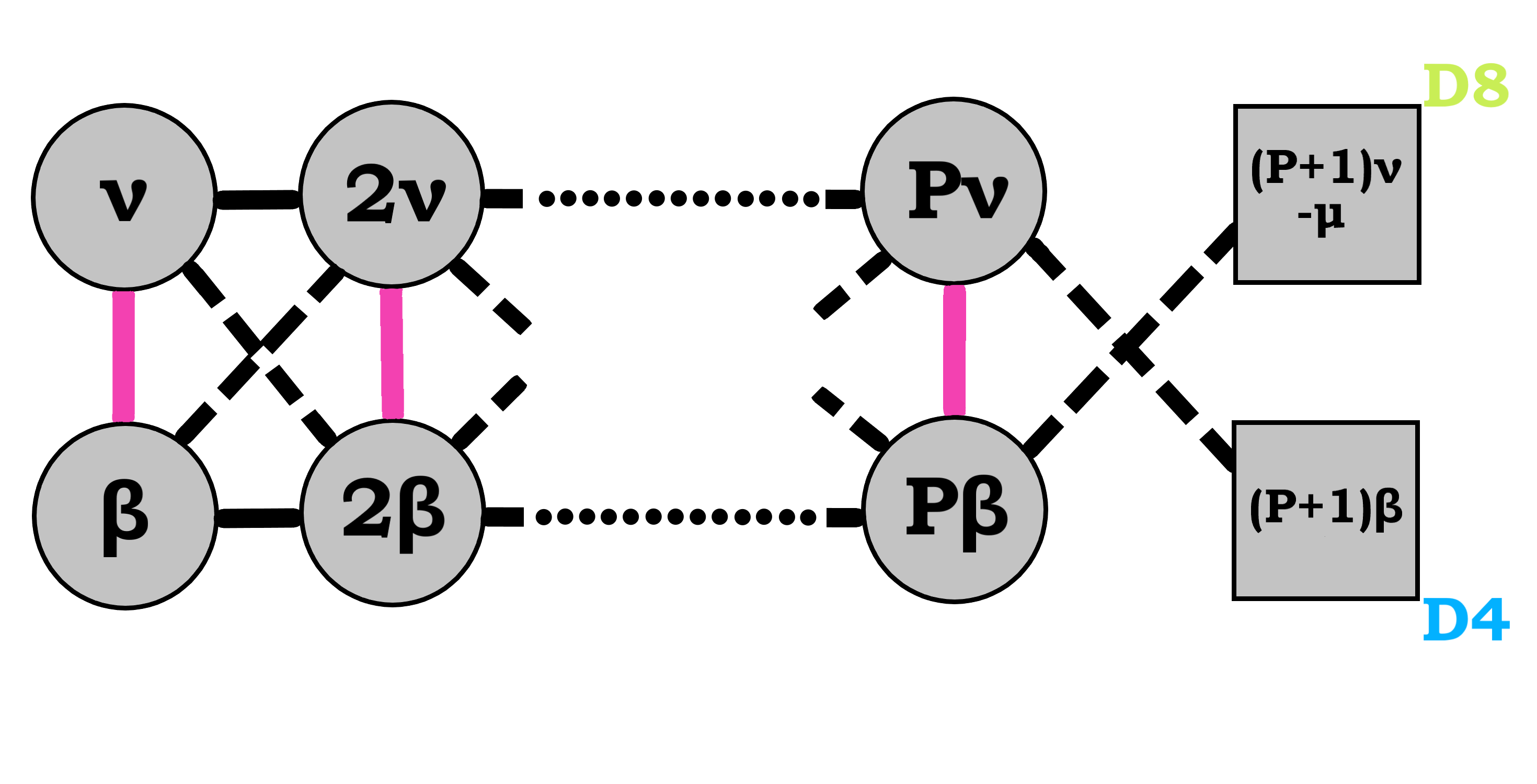}
  \caption{This is the naive quiver dual to the background defined by (\ref{h4example2}), (\ref{h8example2}). In reality, there is one more flavor node, canceling the gauge anomalies for the last D2 gauge node.}
  \label{figure6b}
\end{subfigure}
\caption{A simplified version of the background given in Figure \ref{figure3b} and its dual quiver theory. Here, besides a linear function $u$, $h_4$ starts and closes with a vanishing value, while $h_8$ starts at zero but finishes at a non-zero value.}
\label{figure6}
\end{figure}

Next, let us study the case represented by Figure \ref{figure3b}. Again, we consider Figure \ref{figure6a} instead which falls into the same class of backgrounds but is way simpler. This is the class of backgrounds where $h_8$  does not vanish at the end of the $\rho$-interval while $h_4$ does.

Therefore, according to Figure \ref{figure6a} the defining functions read

\begin{equation}
h_4(\rho)\:=\:\left\lbrace\begin{array}{ccc}
\frac{\beta}{2\pi}\rho &\hspace{1cm}2\pi k\leq\rho\leq2\pi(k+1) &k=0,...,P-1\,,\\
\frac{\beta P}{2\pi}(2\pi(P+1)-\rho) &\hspace{1cm}2\pi P\leq\rho\leq2\pi(P+1) &
\end{array}\right.\label{h4example2}
\end{equation}
\begin{equation}
h_8(\rho)\:=\:\left\lbrace\begin{array}{ccc}
\frac{\nu}{2\pi}\rho &\hspace{1cm}2\pi k\leq\rho\leq2\pi(k+1) &k=0,...,P-1\,,\\
\mu-\frac{\nu P-\mu}{2\pi}(\rho-2\pi(P+1)) &\hspace{1cm}2\pi P\leq\rho\leq2\pi(P+1) &
\end{array}\right.\label{h8example2}
\end{equation}\\

The background defined by these functions is $-$ naively $-$ dual to the quiver theory given by Figure \ref{figure6b}. Again, this quiver cannot be the right one and this can be seen by using the anomaly cancellation condition (\ref{anomalyCond}) on the last D2 gauge node, i.e. the one with gauge rank $P\beta$. For that node the gauge anomalies do not cancel. On the contrary, anomaly cancellation would occur if it connected to a flavor node of rank $\mu$ through a $\mathcal{N}=(0,2)$ Fermi multiplet.

We go on and focus on the dual geometric vicinity of the `anomalous' gauge node, anticipating again to find the necessary portion of D-branes that cancel the gauge anomalies. We find that near the endpoint, $\rho=2\pi(P+1)-x$, for $x\rightarrow0$, the backgrounds reads

\begin{equation}
\dd s^2\;=\; \frac{s_1}{\sqrt{x}}m_1\,\dd s^2_{\mbox{\tiny AdS$_3$}}+\sqrt{x}\left(m_2\,\dd \rho^2+m_3\,\dd s^2_{\mbox{\tiny S$^2$}}+m_4\,\dd s^2_{\mbox{\tiny CY$_2$}}\right)\:,\hspace{1cm}e^{\phi}=m_5\,x^{\frac{1}{4}}\,,\label{exampleIID6}
\end{equation}\\
with $m_i$ real constants, which corresponds to D2-branes on AdS$_3$ and smeared over CY$_2\times$S$^2$. To be exact, this background also corresponds to O2-planes, but strings may live only on D2-branes and, thus, we only consider those to search for global symmetries. Being explicit branes, these D2-branes contribute to the flavor structure of the quiver theory and, in principle, they should cancel the gauge anomalies.

However, we encounter the same problem as with Example I. That is, the Bianchi identities yield that the $h_8$ function only gives rise to D8-branes and certainly not to D2-branes. Therefore, since we do know we should have D8-branes at the endpoint $\rho=\rho_f$ where the $h_8$ function is non-vanishing, while we do not see them, we look up the D8-branes' source terms, that is their Chern-Simons action
\begin{equation}
S_{\mbox{\tiny CS}}^{\mbox{\tiny D8}}\;=\;\mu_8\int\Tr\: C_9^{el}\:+\:C_7^{el}\wedge\mathcal{F}_2\:+\:C_5^{el}\wedge\mathcal{F}_2\wedge\mathcal{F}_2\:+\:C_3^{el}\wedge\mathcal{F}_2\wedge\mathcal{F}_2\wedge\mathcal{F}_2\label{D8WZ}
\end{equation}\\
where the first term sources standard D8-branes and the rest reflect eight-branes as bound states of D6, D4 and D2-branes, respectively. Here, we omitted the coupling to the single D8 transverse mode since there is no object into which this brane could possibly polarize.

Taking into account the RR sector (\ref{RRsector}) near the endpoint $\rho=\rho_f$, we again pick a convenient gauge and deduce

\begin{equation}
\begin{split}
C_7^{el},C_9^{el}\;\;\rightarrow&\;\;const.\,,\hspace{2cm}C_5^{el}\;\propto\;\left(\log(\rho_f-\rho)\right)\,\mbox{vol(AdS$_3$)}\wedge\mbox{vol(S$^2$)}\;\;\rightarrow\;\;-\infty\,,\\[20pt]
&C_3^{el}\;\propto\;\left(\frac{-1}{\rho_f-\rho}\right)\,\mbox{vol(AdS$_3$)}\;\;\rightarrow\;\;-\infty\;.
\end{split}
\end{equation}\\

Since $C_5^{el}$ and $C_5^{el}$ blow up at the boundary, then their corresponding source terms in the Chern-Simons action (\ref{D8WZ}) dominate the game as opposed to the rest. Between those two potentials, $C_3^{el}$ scales infinitely faster as we approach $\rho_f$ and therefore we argue that, at the boundary, the D8-brane gauge field couples to an infinitely strong $C_3^{el}$ RR potential and induces D2-charge on its worldvolume, yielding the analogous background (\ref{exampleIID6}). Additionally, the smearing of those D2-branes can be understood by the coupling of $C_3^{el}$ to $(\wedge\mathcal{F}_2)^3$, in the D8/D2 source term of (\ref{D8WZ}).

We conclude that the D8-branes' gauge field couples to D2-charge through the term

\begin{equation}
S_{\mbox{\tiny CS}}^{\mbox{\tiny D8/D2}}\;=\;\frac{\mu_2}{(2\pi)^3}\int\Tr\:C_3^{el}\wedge\tilde{f}_2\wedge\tilde{f}_2\wedge\tilde{f}_2\label{D8/D2WZterm}
\end{equation}\\
together forming a D8/D2 bound state. The D8 gauge flux on CY$_2\times$ S$^2$ should be quantized as

\begin{equation}
\frac{1}{(2\pi)^3}\int_{\mbox{\tiny CY$_2\times$ S$^2$}}\tilde{f}_2\wedge\tilde{f}_2\wedge\tilde{f}_2\;=\;N_2\hspace{1.5cm}\mbox{for}\hspace{1.5cm}N_2\in\mathbb{Z}
\end{equation}\\
and the D2-branes are explicitly given by the Bianchi identity
\\

\begin{equation}
\dd\hat{f}_6\;=\;\frac{\lambda^3}{3!}\tilde{f}_2\wedge\tilde{f}_2\wedge\dd F_0\;=\;\frac{\lambda^3}{3!}\,N_2\,\mbox{vol(CY$_2$)}\wedge\mbox{vol(S$^2$)}\wedge(h_8''\,\dd\rho)
\end{equation}\\

Hence, we conclude that every eight-brane on the boundary should exist exclusively in a D8/D2 bound state, sourced by

\begin{equation}
S_{\mbox{\tiny CS}}^{\mbox{\tiny 1$\times$D8/D2}}\;=\;N_2\left(\mu_2\int C_3^{el}\right)\label{oneD8/D2WZterm}
\end{equation}\\
that is each D8-brane contains $N_2$ units of D2-charge.

Nonetheless, there is no just one D8-brane (with an Abelian gauge field) but there should be multiple \textit{coincident} D8-branes at the boundary. The number of these branes is given by the Bianchi identity

\begin{equation}
\dd\hat{F}_0\;=\;h_8''\,\dd\rho
\end{equation}\\
where, following the same procedure for $h_8'$ as in Example I with $h_4'$, we find that at the boundary $\rho=\rho_f$ they amount to

\begin{equation}
N_8\Big|_{\rho=\rho_f}\;=\;\mu
\end{equation}\\
Since those D8-branes are coincident and thus their gauge field is non-Abelian, a U$(\mu)$ gauge theory arises that is realized as a global symmetry in the dual quiver theory and which should cancel the apparent gauge anomalies there.

\begin{figure}[t!]
    \centering
    %\subfloat[label 1]
    {{\includegraphics[width=10.8cm]{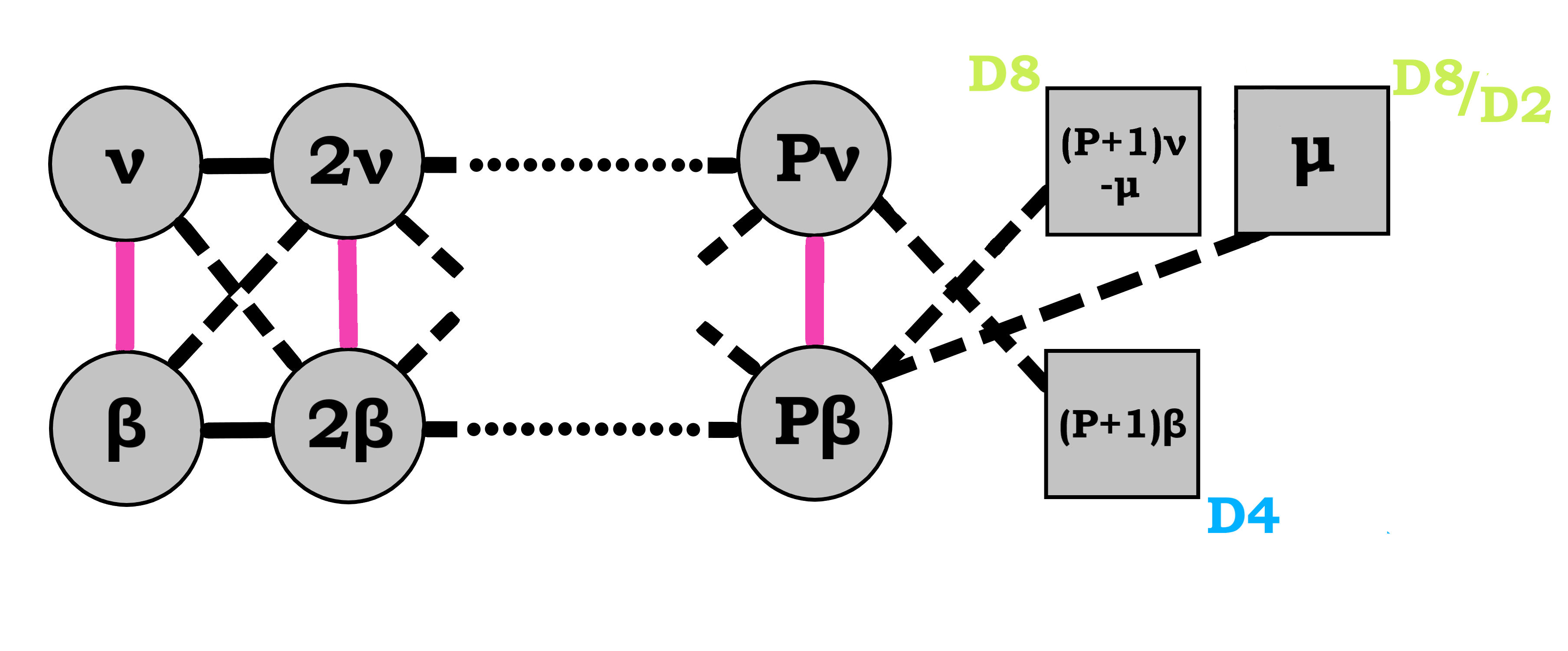} }}%
   % \qquad
    %\subfloat[label 2]
    %{{\includegraphics[width=4.5cm]{quiver_1_alpha_p.jpg} }}%
    %\qquad
    %\subfloat[label 3]
    %{{\includegraphics[width=4.5cm]{quiver_1_alpha_pp.jpg} }}
   % \caption{2 Figures side by side}%
\caption{This is the actual quiver dual to the background defined by (\ref{h4example2}), (\ref{h8example2}). Here, the extra D2  and D6 flavor nodes cancel the gauge anomalies for the first D6 and the last D2 gauge nodes.}
\label{figure13}
\end{figure}

Indeed, the D8-branes, as D8/D2 bound states, feed with flavor the end of the D2 color chain of the quiver through a $\mathcal{N}=(0,2)$ Fermi multiplet, as usual. As expected, using the anomaly cancellation condition (\ref{anomalyCond}), they give exactly the flavor needed in order to cancel the gauge anomalies of the last D2 node. This is all visualized in Figure \ref{figure13}, where the quiver theory is now consistent.\\

\subsection{Constant $u(\rho)$}\label{Section3.2}
\begin{figure}
\centering
\begin{subfigure}[b]{0.48\textwidth}
 \centering
  \includegraphics[width=1\linewidth]{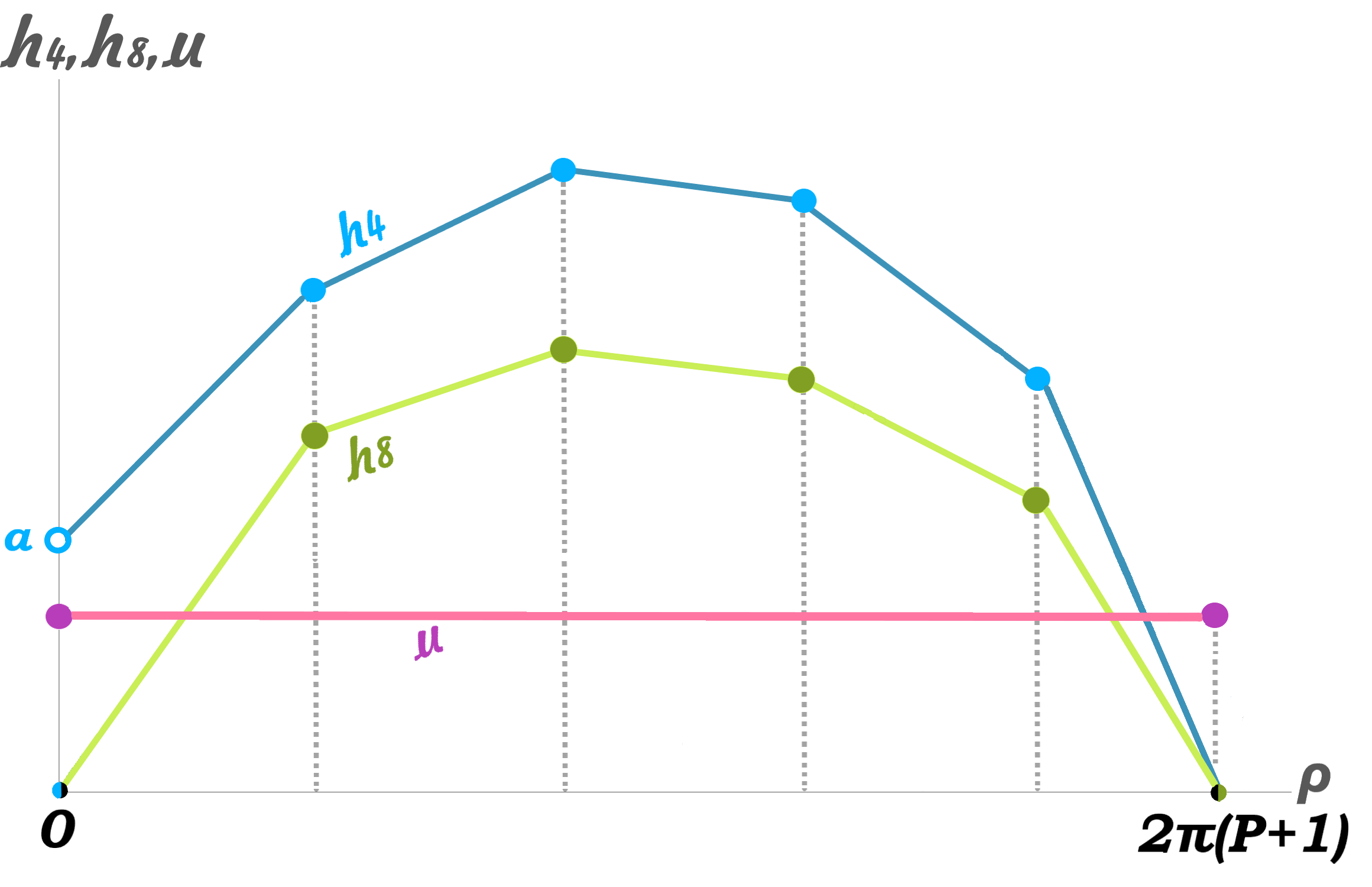}
  \caption{A background with constant $u$ and a non-vanishing $h_4$ at the beginning $\rho=0$.}
  \label{figure8a}
\end{subfigure}
\hspace{1em}%
\begin{subfigure}[b]{.48\textwidth}
  \centering
  \includegraphics[width=1\linewidth]{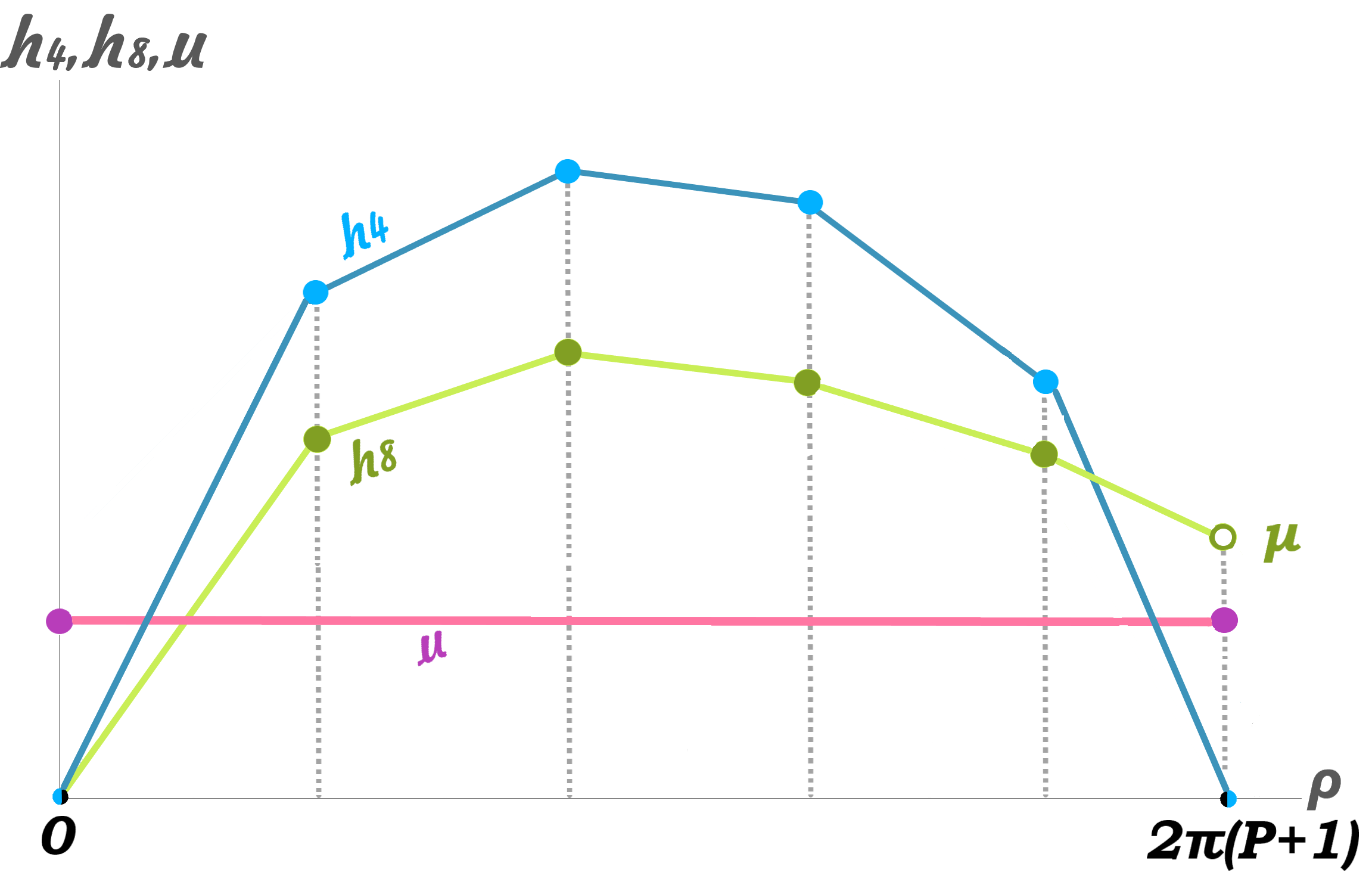}
  \caption{A background with constant $u$ and a non-vanishing $h_8$ at the endpoint $\rho=\rho_f$.}
  \label{figure8b}
\end{subfigure}
\caption{The representative backgrounds defined by a constant $u(\rho)$ and a non-vanishing $h_4$ or $h_8$ at either endpoint. The roles of $h_4$ and $h_8$ may be exchanged in (a) and (b).}
\label{constantu}
\end{figure}

The class of supergravity backgrounds with constant function $u(\rho)$ is analogous but, at the same time, dissimilar to the linear case. The representative kinds of backgrounds in this class are the ones presented in Figures \ref{constantu}, distinguished by their constant $u(\rho)$ curve. Instead of going through both examples again, we now combine them into one that includes all the interesting behavior. That is, at the beginning of the $\rho$-dimension $h_4$ does not vanish while $h_8$ does, the opposite being true at the other endpoint. Of course, we again realize simplified versions of these cases as in the previous examples and, depending on the behavior of the defining functions at each endpoint, the precise form of $h_4$ and $h_8$ can be read off from (\ref{h4example1}),(\ref{h8example1}) and (\ref{h4example2}),(\ref{h8example2}). Accordingly, for this new background, we seek for U$(\alpha)$ and U$(\mu)$ flavor symmetries at $\rho=0$ and $\rho=\rho_f$ respectively, in order to cure the apparent gauge anomalies at the dual edge-nodes of the quiver chain.

\subsubsection*{At the beginning of the $\rho$-dimension}
The background we consider begins on its $\rho$-dimension, for $\rho=x$ while $x\rightarrow0$, with a vanishing $h_8$ but a non-vanishing $h_4$ function, giving
\begin{equation}
\dd s^2\;=\;\frac{1}{\sqrt{x}}\left(m_1\,\dd s^2_{\mbox{\tiny AdS$_3$}}+m_2\,\dd s^2_{\mbox{\tiny S$^2$}}+m_3\,\dd s^2_{\mbox{\tiny CY$_2$}}\right)+m_4\sqrt{x}\,\dd x^2\:,\hspace{1cm}e^{\phi}=m_5\,x^{-\frac{5}{4}}\,,\label{exampleIIu=cD8}
\end{equation}\\
that corresponds to D8-branes on AdS$_3\times$S$^2\times$CY$_2$, which again seems odd since $h_4$ only gives D4-branes. Our experience gained from the precious sections drives us to study the full Chern-Simons source action of $N_4$ D4-branes, including the coupling of the transverse string modes to the higher dimensional RR fields, as
\begin{equation}
S_{\mbox{\tiny CS}}^{D4}\;=\;\mu_4\int\Tr\:C_5^{el}+C_3^{el}\wedge\mathcal{F}_2+i\lambda(\imath_\Phi\imath_\Phi)C_7^{el}-\lambda^2(\imath_\Phi\imath_\Phi)^2\left(C_9^{el}+C_7^{el}\wedge\mathcal{F}_2+\ldots\right)\label{D4WZu=c}
\end{equation}\\
where the first term represents standard D4-branes and the second D4/D2 bound states, while the rest reflect polarized D4-branes into higher dimensional ones. Considering the RR sector (\ref{RRsector}) near the beginning $\rho=0$, we deduce

\begin{equation}
C_3^{el},C_7^{el}\;\;\rightarrow\;\;0\;\;,\hspace{2cm}C_5^{el}\;\;\rightarrow\;\;const.\;\;,\hspace{2cm}C_9^{el}\;\;\rightarrow\;\;-\infty\,,
\end{equation}\\
at the vicinity of that boundary, where again a convenient gauge was chosen.

Therefore, at $\rho\rightarrow0$, only the first and fourth term survive in (\ref{D4WZu=c}), which stand for standard D4-branes and D4/D8 bound states, respectively. Since the potential $C_9^{el}$ blows up, without any competition this time, the fourth term in the above action dominates the first and this is why the background metric and dilaton behave according to (\ref{exampleIIu=cD8}). That is, the D4-branes couple to an infinitely strong RR potential $C_9^{el}$ and condense out into an eight-brane, forming a D8/D4 bound state while giving a D8-brane background on that boundary. Of course, the non-vanishing $C_5^{el}$ is vital for the very existence of those constituent D4-branes. As it is the case with Example I and (\ref{D4WZ}), both the coupling to the transverse scalars and the string length order in the Chern-Simons action (\ref{D4WZu=c}) would make here a more detailed treatment instructive, a calculation that is held in Appendix \ref{AppendixD8/D4}. 

Casting the usual trick on $h_4'$, we count $\alpha$ D4-branes on $\rho=0$, on which open strings end and make up a U$(\alpha)$ gauge theory. The polarization that takes place over CY$_2$ should raise the question whether the D4-branes are enough in number, throughout the bound state, to support massless string modes and thus a unitary gauge theory. As restated though, we do know that the D4-branes are a lot since $\alpha$ must be also large in the supergravity limit, by construction. Therefore U$(\alpha)$ is the flavor group we anticipated for the beginning node of the quiver chain, canceling exactly the gauge anomalies there through a $\mathcal{N}=(0,2)$ Fermi multiplet.

\subsubsection*{At the end of the $\rho$-dimension}
Focusing on the other endpoint, $\rho=2\pi(P+1)-x$ while $x\rightarrow0$, the same background ends on its $\rho$-dimension with a vanishing $h_4$ but a non-vanishing $h_8$, giving

\begin{equation}
\dd s^2\;=\; \frac{1}{\sqrt{x}}\left(s_1\,\dd s^2_{\mbox{\tiny AdS$_3$}}+s_2\,\dd s^2_{\mbox{\tiny S$^2$}}\right)+\sqrt{x}\left(s_3\,\dd x^2+s_4\,\dd s^2_{\mbox{\tiny CY$_2$}}\right)\:,\hspace{1cm}e^{\phi}=s_5\,x^{-\frac{1}{4}}\,,
\end{equation}\\
which corresponds to D4-branes smeared over CY$_2$. While this seems odd since $h_8$ only produces D8-branes, our wisdom off the previous section guides us to study the source terms

\begin{equation}
S_{\mbox{\tiny CS}}^{\mbox{\tiny D8}}\;=\;\mu_8\int C_9^{el}\:+\:C_7^{el}\wedge\mathcal{F}_2\:+\:C_5^{el}\wedge\mathcal{F}_2\wedge\mathcal{F}_2\:+\:C_3^{el}\wedge\mathcal{F}_2\wedge\mathcal{F}_2\wedge\mathcal{F}_2\label{D8WZu=c}
\end{equation}\\
where the first term sources a standard D8-brane and the rest reflect a D8-brane in a bound state with D6, D4 and D2-branes, respectively.

Studying the RR fluxes (\ref{RRsector}) at $\rho\rightarrow\rho_f$ for a constant function $u$ again, the potentials behave as

\begin{equation}
C_7^{el}\;\;\rightarrow\;\;0\;\;,\hspace{2cm}C_3^{el},C_9^{el}\;\;\rightarrow\;\;const.\;\;,\hspace{2cm}C_5^{el}\;\;\rightarrow\;\;-\infty\,,
\end{equation}\\
where we again chose a convenient gauge. The fact that $C_7^{el}$ vanishes excludes the D8/D6 bound state whatsoever. Between the rest of the terms in (\ref{D8WZu=c}), the one that couples to $C_5^{el}$ dominates since it is this potential that blows up at the vicinity of that endpoint.

We conclude that the D8-brane gauge field couples to D4-charge through the term

\begin{equation}
S_{\mbox{\tiny CS}}^{\mbox{\tiny D8/D4}}\;=\;\frac{\mu_4}{4\pi^2}\int\Tr\:C_5^{el}\wedge\tilde{f}_2\wedge\tilde{f}_2\label{D8/D4WZterm}
\end{equation}\\
together forming a D8/D4 bound state. The fact that $C_5^{el}$ is infinitely strong makes the source term (\ref{D8/D4WZterm}) dominant in (\ref{D8WZu=c}) and this is why the eight-branes are geometrically realized as smeared D4-branes. The D8 gauge flux on CY$_2$ should be quantized as

\begin{equation}
\frac{1}{4\pi^2}\int_{\mbox{\tiny CY$_2$}}\tilde{f}_2\wedge\tilde{f}_2\;=\;N_4\hspace{1.5cm}\mbox{for}\hspace{1.5cm}N_4\in\mathbb{Z}
\end{equation}\\
and the D4-branes are explicitly given by the Bianchi identity

\begin{equation}
\dd\hat{f}_4\;=\;\frac{\lambda^2}{2}\tilde{f}_2\wedge\tilde{f}_2\wedge\dd F_0\;=\;\frac{\lambda^2}{2}\,N_4\,\mbox{vol(CY$_2$)}\wedge(h_8''\,\dd\rho)
\end{equation}\\
Hence, we conclude that every eight-brane on the boundary should exist exclusively in a D8/D4 bound state, sourced by

\begin{equation}
S_{\mbox{\tiny CS}}^{\mbox{\tiny 1$\times$D8/D4}}\;=\;N_4\left(\mu_4\int C_5^{el}\right)\label{oneD8/D2WZterm}
\end{equation}\\
that is each D8-brane contains $N_4$ units of D4-charge.

Nonetheless, there is no just one D8-brane but there should be multiple \textit{coincident} D8-branes at the boundary. The number of these branes, same as in the last section with Example II, is given by $N_8=\mu$. Since those D8-branes are coincident and thus their gauge field is non-Abelian, a U$(\mu)$ gauge theory arises that is realized as a global symmetry in the dual field theory and which cancels exactly the gauge anomalies in the end of the quiver chain through a $\mathcal{N}=(0,2)$ Fermi multiplet.

Note that the smeared D4 and the D8-branes in this section are backgrounds equivalent to smeared O4 and O8-planes, respectively. Of course, strings may only live on the former which is why we only consider those to find the desired flavor symmetries.\\

As a last remark on the whole section, let us clarify a few details about the RR potentials. Firstly, the fact that we chose a particular gauge does not change any of the results. Indeed, by studying the RR fluxes we realize that had we picked any other gauge choice would have made no difference; the qualitative relationship between the $C_p$ forms (which one is stronger at the endpoints) would have stayed the same. Secondly, one may wonder whether such objects blowing up test the supergravity approximation. However, as argued in \cite{Lozano:2019zvg}, singularities are bound to exist when D-branes do, while they are not dangerous as long as they are regulated and stay far apart from each other (here, along the $\rho$-dimension). This is exactly the case with the Ricci scalar (which diverges at the positions of localized sources) and with the RR potentials, as long as $\beta_k,\nu_k,P$ are large. Indeed, large $\beta_k,\nu_k$ control all divergences, while large $P$ keeps the singularities far apart (for the backgrounds we considered, RR potentials only blow up at the endpoints, anyway). Nonetheless, we believe that the particular divergence of some of the RR potentials at the endpoints is an artifact of the functions $h_4,h_8$ being defined on a closed interval; this was the case when we counted D-branes at those endpoints, where we had to go around the fact that $h'_4,h'_8$ are not well-defined there. The essence of those infinities in our context is that some potentials are profoundly stronger than others.\\

Aside from curing a problem and better realizing the way the dual field theory works, this section has an additional value. Since the discovery of particular flavor branes was the exact thing that made the quiver theory consistent, this calculation provides an additional validity check of the whole field theoretical structure. Further validation of the quantum quiver structure is  especially important here, since the matter content of these quiver theories is by no means trivial. This is the subject of the following section.\\

\section{Adding matter in the quiver field theory}\label{Section4}
The quantum quiver theory dual to the AdS$_3$ supergravity vacua we consider was presented in Section \ref{subsectionQUIVERtheory}. In \cite{Lozano:2019zvg} these linear quiver theories were thoroughly analyzed and tested, while our previous section suits as further validation. Nevertheless, there is more to their story to tell. That is they are ultimately characterized by additional structure.

Let us address the problem in a constructive way. In a Hanany-Witten brane set-up, we have all possible kinds of oscillating strings stretched between the branes. In the dual quiver theory, these kinds of strings correspond to supersymmetric multiplets that bind the gauge theories (gauge nodes) together and constitute the matter content of the overall field theory. Thus, when we try to build the correct dual field theory of a particular kind of brane set-up, the problem boils down to finding all the possible matter content.

Establishing the quiver theory introduced \cite{Lozano:2019jza,Lozano:2019zvg,Lozano:2019ywa} as a well tested structure, we realize that there are two kinds of superfield connection missing. These are the multiplets connecting D2 gauge with D4 flavor nodes and the ones connecting D6 gauge with D8 flavor nodes, respectively representing D2-D4 and D6-D8 strings. Instead of quantizing, we may just ask what multiplets can possibly fill this gap. The problem gets quickly simplified, since we know we do not want to consider additional $\mathcal{N}=(0,4)$ hyper multiplets nor $\mathcal{N}=(0,2)$ Fermi multiplets. This is because their presence would spoil the fragile balance of the gauge anomaly cancellation once and for all, a balance that was further confirmed to holographically hold by the last section. Therefore, we should only consider $\mathcal{N}=(4,4)$ hyper multiplets.

Nonetheless, our unique choice should be in harmony with the central charge of the field theory. In particular, since the central charge was found in \cite{Lozano:2019zvg} to be holographically correct for the (original) quiver theory, then the new matter content we want to add should change nothing and be entirely invisible to it. Indeed, this is exactly the case. The central charge of the quiver field theory reads
\begin{equation}
c=6\left(n_{hyp}-n_{vec}\right)=6\left(\sum_{j=1}^P\left(\alpha_j\mu_j-\alpha_j^2-\mu_j^2+2\right)+\sum_{j=1}^{P-1}\left(\alpha_j\alpha_{j+1}+\mu_j\mu_{j+1}\right)\right)\label{CentralChargeQuiver}
\end{equation}\\
which means that it is sensitive to the number of the hyper multiplets. This may sound discouraging wrt adding new $\mathcal{N}=(4,4)$ hyper multiplets, since we want to leave the central charge intact, but it is not. This is because we work in the supergravity limit, i.e. for $P\rightarrow\infty$, which means that we are eligible to add new hyper multiplets as long as their number is sub-leading in $P$ wrt to the old ones.

\begin{figure}[t!]
    \centering
    %\subfloat[label 1]
    {{\includegraphics[width=6cm]{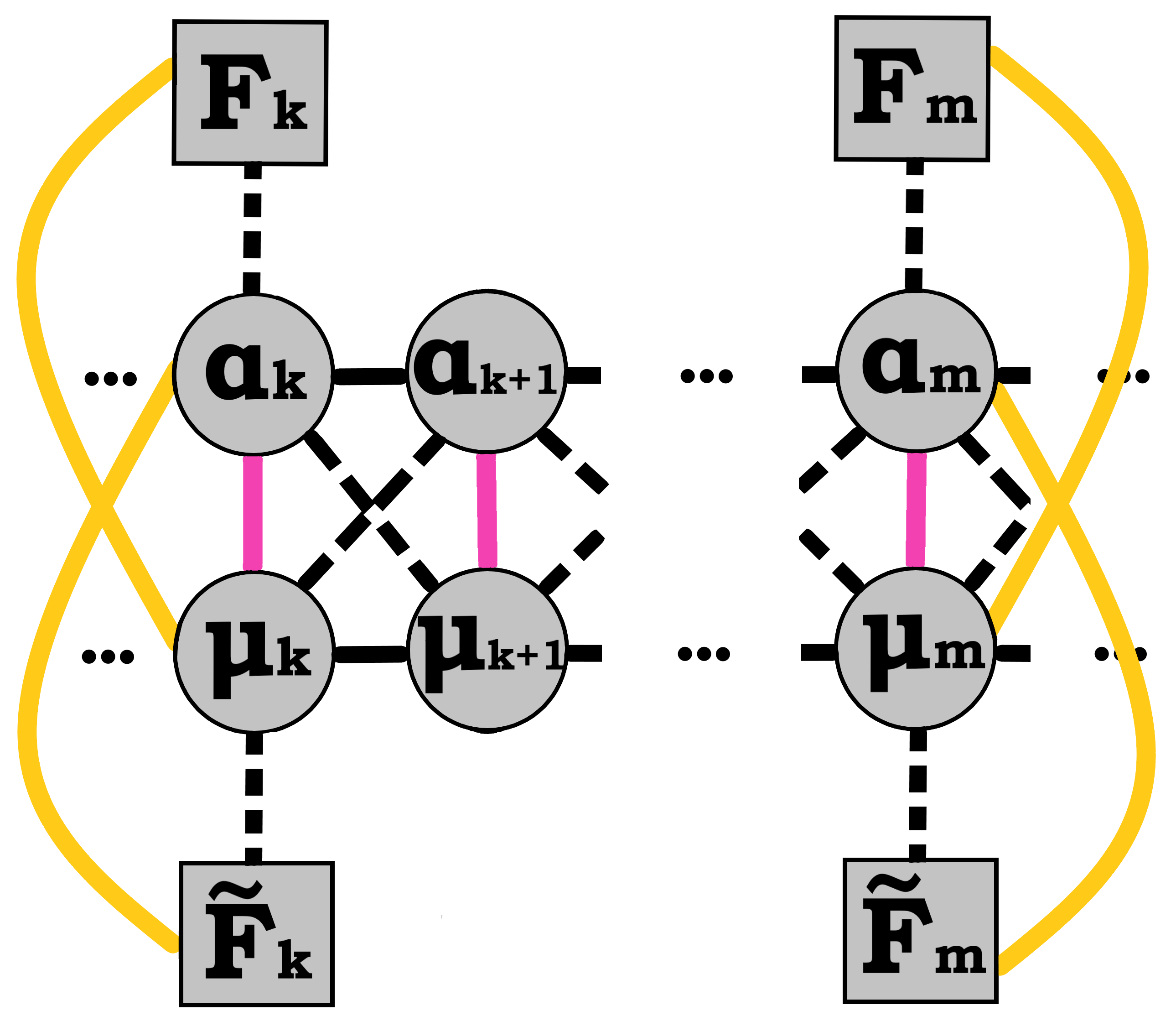} }}%
   % \qquad
    %\subfloat[label 2]
    %{{\includegraphics[width=4.5cm]{quiver_1_alpha_p.jpg} }}%
    %\qquad
    %\subfloat[label 3]
    %{{\includegraphics[width=4.5cm]{quiver_1_alpha_pp.jpg} }}
   % \caption{2 Figures side by side}%
\caption{This is the new dual quiver theory, with additional $\mathcal{N}=(4,4)$ hyper multiplets binding the D4 and D8 flavor nodes with the D2 and D6 gauge nodes, respectively. The already existing $\mathcal{N}=(4,4)$ hyper multiplets are represented with black solid lines, while the new additional ones with orange solid lines.}
\label{figure18}
\end{figure}

In the supergravity limit the sources (flavor nodes) should exist far apart along the linear quiver, which means that the new hyper multiplets escorting them are much less than the old ones that exist between the flavor positions (connecting the gauge nodes). The proposed, enhanced quiver theory is visualized in Figure \ref{figure18}.

In order to prove that the new hyper multiplets are always of lower order in $P$ than the old ones, we expand the already existing ones as

\begin{equation}
n_{hyp}\;=\;\sum_{j=1}^P\left(\sum_{k=0}^{j-1}\beta_k\,\cdot\,\sum_{l=0}^{j-1}\nu_l\right)+\sum_{j=1}^{P-1}\left[\left(\sum_{k=0}^{j-1}\beta_k\,\cdot\,\sum_{l=0}^{j}\beta_l\right)+\left(\sum_{k=0}^{j-1}\nu_k\,\cdot\,\sum_{l=0}^{j}\nu_l\right)\right]\label{nhypOLD}
\end{equation}\\
while the new ones, $n_{hyp}^\star$, read
\begin{equation}
\begin{split}
n_{hyp}^\star\;&=\;\sum_{j=i_1}^{i_M}\alpha_j\tilde{F}_{j-1}+\sum_{j=i_1}^{i_N}\mu_jF_{j-1}\\
&=\;\sum_{j=i_1}^{i_M}\left(\sum_{k=0}^{j-1}\beta_k\left(\beta_{j-1}-\beta_j\right)\right)+\sum_{j=i_1}^{i_N}\left(\sum_{k=0}^{j-1}\nu_k\left(\nu_{j-1}-\nu_j\right)\right)
\end{split}\label{nhypNEW}
\end{equation}
where $j=i_1,...,i_{M,N}$ are the $M,N$ intervals with sources for the D4 and D8 branes, respectively. The fact that in the supergravity limit the sources (flavor nodes) should exist far apart along the linear quiver means $M,N\ll P$.

In order to compare $n_{hyp}$ and $n_{hyp}^\star$ we can just focus into similar terms between them. These are, for instance, the second term of (\ref{nhypOLD}) and the first of (\ref{nhypNEW}). For them, we observe that their first summation is to $P-1$ and $i_M$, respectively. Since $M,N\ll P$, this means that the former is of order $P$ while the latter is not. Focusing on the inner summations of the same terms, we realize that their summing products are of the same order, whatever that is. Therefore, overall, $n_{hyp}$ is always an order higher in $P$ than $n_{hyp}^\star$, which makes the latter invisible in the central charge for $P\rightarrow\infty$.

The whole situation would be immediately cleared out if we quantized the system of D-branes. What is more, quantizing the D2-D4 and D6-D8 systems in flat space seems to indeed reproduce the new $\mathcal{N}=(4,4)$ hypermultiplets that we just proposed to exist. However, this particular Hanany-Witten set-up is assumed to live in CY$_2$ dimensions as well, which makes the standard quantization techniques obscure in the case at hand and, therefore, such a study remains on the sidelines at this point.

Another link that we intentionally left out is the multiplet corresponding to superstrings between D4 and D8 flavor branes. Those superfields transform in the bi-fundamental representation of two flavor groups, they do not couple to vector superfields and, thus, are not gauged. Hence, they decouple from the quiver gauge theory.\\

Truth be told, there is another path through which we might have imagined that the additional matter is an essential ingredient to our theory. This argument too surfaces from the supergravity side of the duality, but in order to illustrate it we need to consider a particular state of the bosonic string. This is what we deal with in the following section.

\section{The meson string}\label{Section5}
Having worked out even the most exotic parts of the correspondence between the massive IIA vacua and the dual quantum field theory, we are certainly in desire of testing their holographic performance. In that vein, we look for a simple object to construct, starting off with the supergravity side of the story.

\subsection{A BPS state}

The most accessible state in our theory of gravity is a semiclassical string stretching between D-branes. That is, we consider a meson string soliton $\mathcal{M}_{k,m}$ on the supergravity background, that extends between stacks of flavor branes at $\rho=2\pi k$ and $\rho=2\pi m$, respectively, and which is a point on the rest of the dimensions sitting at the center $r=0$ of AdS$_3$. An analogous calculation was performed in \cite{Bergman:2020bvi}.

Therefore, we allow a string embedding with $\tau=t, \sigma=\rho$, whose mass is essentially its length
\begin{equation}
M_{\mbox{\tiny$\mathcal{M}$}}\;=\;\frac{1}{2\pi}\int\dd\sigma\sqrt{-\det g_{ab}}\;=\;\frac{1}{2\pi}\int^{2\pi m}_{2\pi k}\dd\rho\sqrt{-\det g_{ab}}\;=\;m-k\label{BPSmass}
\end{equation}\\
where $g_{ab}$ is the worldsheet pullback of the metric in (\ref{GeneralAdS3background}). If $F_k$ and $F_m$ are the number of D-branes in the respective stacks where the string endpoints end, then this configuration transforms in the bi-fundamental representation of SU$(F_k)$ $\times$ SU$(F_m)$.

Since we are always interested in states that preserve some supersymmetry, we may upgrade the above configuration to a BPS state just by considering the suspended string to fluctuate on the two-sphere, whose SU$(2)$ isometry corresponds to the dual R-symmetry. This is done by including $\phi=\omega\tau$ in the above configuration, where we let this fluctuation to be small $-$ i.e. $\omega\ll1$ $-$ so that the embedding simplifies still into the expression (\ref{BPSmass}).

Picking a U$(1)_R$ inside SU$(2)_R$, we now seek the R-charge of the above state. Since the generator of the U$(1)$ on the two-sphere is associated to the 1-form $\cos\theta\,\dd\phi$, then we look for the string coupling terms
\begin{equation}
S_R\;\propto\;\int\cos\theta\,\dd\phi
\end{equation}\\
As far as the R-charge is concerned, it may be read off the source terms of the form $\int J_R A_1=Q_R\int A_1$, with $A_1=\cos\theta\,\dd\phi$. The relevant term in the worldsheet action is

\begin{equation}
S_{\mbox{\tiny$\mathcal{M}$}}=\frac{1}{2\pi}\int_\Sigma B_2
\end{equation}\\
where $\Sigma=[2\pi k,2\pi m]\times\mathbb{R}$. Ultimately, after some manipulation given in Appendix \ref{appendixRcharge}, this term may be actually seen as the source term

\begin{equation}
S_{\mbox{\tiny$\mathcal{M}$}}\;=\;\left(m-k\right)\int_{\mathbb{R}}\cos\theta\,\dd\phi
\end{equation}
which yields an R-charge

\begin{equation}
Q_R\;=\;m-k\label{BPScharge}
\end{equation}\\
Comparing this with the string mass in (\ref{BPSmass}), we conclude that this is indeed a BPS state.

\subsection{An ultraviolet operator}
Now, we want to look for the operator dual to this BPS state. To this end $-$ since the IR SCFT is completely unknown $-$ we consider the UV quiver theory on the $\rho$-interval $[2\pi k,2\pi m]$ and pick the appropriate field excitations inside the supersymmetric multiplets.

Since we are dealing with a purely bosonic state, we are immediately led to consider the complex scalars $\phi_i$ inside the $\mathcal{N}=(0,2)$ chiral multiplets $\Phi_i$, since these are the obvious on-shell bosonic degrees of freedom in our theory. In particular, we choose to excite one scalar in each of the $(m-k)+2$ $\mathcal{N}=(4,4)$ hypermultiplets that connect two flavor nodes; this makes a perfect fit with the fact that string fluctuations transverse to the worldvolumes of branes are also scalar modes wrt these worldvolume theories.  It also illustrates why we need the additional $\mathcal{N}=(4,4)$ matter, as promised in the beginning of this section; if it was not for these new hypermultiplets, there would be no way to build a string of bosonic field excitations that connect two flavor nodes. And such a dual bosonic connection must somehow exist, given that the meson string we consider is a legitimate BPS state.

Shortly, however, we spot a problem. As illustrated in Appendix \ref{appendixRanomaly}, the $\phi_i$ scalars inside any of the $\mathcal{N}=(0,4)$ hypermultiplets are uncharged under R-symmetry, while we do need an R-charge $-$ according to (\ref{BPScharge}), proportional to $(m-k)$ $-$ for our proposed operator. In fact, the only scalars that are charged under the U$(1)_R$ subgroup of the R-symmetry are the ones in the $\mathcal{N}=(0,4)$ twisted hypermultiplets $(\Sigma_i,\tilde{\Sigma}_i)$, inside the $\mathcal{N}=(4,4)$ vector superfields of the gauge nodes. This leads us to consider these scalars, let us call them $\sigma_i$, as well. The inclusion of these scalar fields is also somewhat compelling, since these are the ones that let the $\phi_i$ scalars interactively talk to each other; this realizes an interactive continuance among the string of fields in the operator, holographically analogous to the compactness of the string. These supersymmetric interactions will become apparent shortly.

\begin{figure}[t!]
    \centering
    %\subfloat[label 1]
    {{\includegraphics[width=6cm]{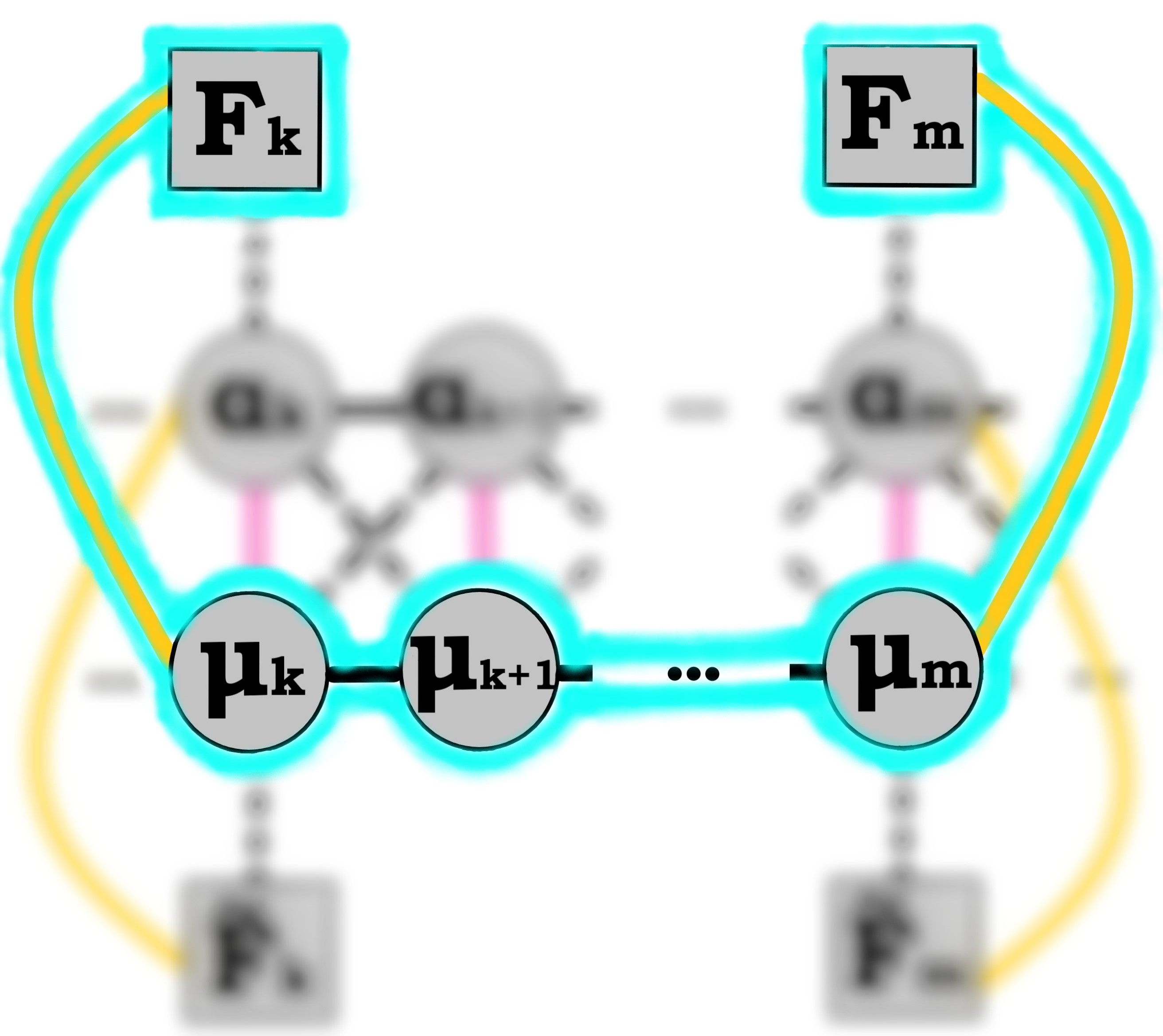} }}%
   % \qquad
    %\subfloat[label 2]
    %{{\includegraphics[width=4.5cm]{quiver_1_alpha_p.jpg} }}%
    %\qquad
    %\subfloat[label 3]
    %{{\includegraphics[width=4.5cm]{quiver_1_alpha_pp.jpg} }}
   % \caption{2 Figures side by side}%
\caption{The meson operator $\mathcal{M}$ consists of the supersymmetric multiplets that are highlighted with blue, while the rest of the quiver structure is left blurred. If $k$ and $m$ are the positions of the flavor nodes along the quiver chain, then this operator runs over $m-k+2$ $\mathcal{N}=(4,4)$ hypermultiplets and $m-k+1$ $\mathcal{N}=(4,4)$ vector multiplets. Such an operator may also connect D4 with D8 flavors, by jumping through $\mathcal{N}=(0,4)$ hypermultiplets.}
\label{figure21}
\end{figure}

All in all, choosing a $\sigma_i$ excitation as well in each gauge node between the $\mathcal{N}=(4,4)$ hypermultiplets, we acquire the meson operator
\begin{equation}
\mathcal{M}_{k,m}\;=\;\pi_k\left(\prod_{i=k}^{m-1}\sigma_i\phi_i\right)\sigma_m\tilde{\pi}_m\label{MesonOperator}
\end{equation}\\
which transforms in the bifundamental representation of SU$(F_k)$ $\times$ SU$(F_m)$, with $F_k$ and $F_m$ the ranks of the flavor groups in the corresponding positions of the quiver chain. Here we named $\pi_i$ the scalars inside the end-point hypermultiplets connecting to the flavor nodes and also chose them to be in conjugate representations of each gauge group. Such an operator has two $\pi_i$'s, $(m-k)$ $\phi_i$'s and $(m-k+1)$ $\sigma_i$'s, which in the supergravity limit $-$ where sources are far apart $-$ account for $2(m-k)$ complex scalars. Since only half of those (the $\sigma_i$'s) are R-charged, this is the desired R-charge considering the BPS string charge (\ref{BPScharge}). For clarity, the operator is highlighted in Figure \ref{figure21}.

The only quantities left to compare are the mass (\ref{BPSmass}) of the BPS state and the conformal dimension of the operator $\mathcal{M}_{k,m}$. At this point, of course, we may have an actual problem; scalar fields in two dimensions have mass dimension zero. At least classically. At first sight, this degrades our proposal for the operator which seems to have a vanishing scaling dimension. However, before rushing into conclusions, we remind ourselves that we have actually considered the UV operator and not the actual IR situation; it is the IR operator the one that should necessarily acquire the appropriate scaling dimension. Therefore, if the choice of operator is correct, our only way out is the possibility of the operator acquiring an anomalous dimension through quantum effects. Whatever the case is with the IR SCFT, such quantum effects should be present in the UV Lagrangian, pointing towards an anomalous dimension $\gamma(g)$ that scales with energy.

On the other hand, studying quantum corrections is obscure in our case. This is exactly because it is the UV theory that we use to organize fields into an operator; therefore even if we assume a completely anomalous dimension $\Delta_{\mathcal{M}}=\gamma(g)$, our SCFT is assumed to be strongly coupled which discredits any perturbative calculation. To be exact, it is the non-integrability of our AdS$_3$ backgrounds \cite{Filippas:2019ihy} that prohibits surfing along the range of the coupling constant, as it is possible with e.g. the work of BMN \cite{Berenstein:2002jq} in the AdS$_5$ $\times$ S$^5$ correspondence. Regardless, the possibility itself of a non-perturbative anomalous dimension requires certain interactions to be there, between the fields of interest; finding whether those exist is essential to our proposal. Interestingly, such interactions indeed exist.

The interactions between the $\phi_i$'s of the hypermultiplets and the $\sigma_i$'s of the twisted hypermultiplets have actually already appeared in our study of the Fermi multiplet interactions. As seen in Section \ref{subsectionSuperpotential}, Fermi multiplets defined by $\bar{\mathcal{D}}_+\Gamma_a=E_a(\Phi_i,\Sigma_i)$ give a potential $\abs{E_a(\phi_i,\sigma_i)}^2$, which for our interactive chain of multiplets exhibits quite a few components. From those, the ones that couple $\phi_i$'s and $\sigma_i$'s are the
\begin{equation}
E_{\Gamma_i}(\phi_i,\sigma_i)\;=\;\sigma_i\phi_i\label{MesonEterm}
\end{equation}
or $E_{\tilde{\Gamma}_i}=-\tilde{\phi}_i\sigma_i$, depending on which scalar field we excite inside a certain hypermultiplet. Accordingly, if we choose to excite $\tilde{\sigma}_i$ inside a twisted hypermultiplet, instead of its twin $\sigma_i$, then these scalars couple through the superpotential term $\abs{J_a(\phi_i,\sigma_i)}^2$ and, in particular, through the components
\begin{equation}
J_{\tilde{\Gamma}_i}(\phi_i,\tilde{\sigma}_i)\;=\;\tilde{\sigma}_i\phi_i\label{MesonJterm}
\end{equation}
or $J_{\Gamma_i}=\tilde{\phi}_i\tilde{\sigma}_i$.

These are all the interactions present between the different scalars we choose to excite and which furnish our operator (\ref{MesonOperator}) with quantum effects. We presume that those are capable of correcting it non-perturbatively to the desired conformal dimension $\Delta_{\mathcal{M}}=\gamma(g)=m-k$.\\

\subsection{Dual mass}
While the scaling dimension of the meson operator stands as a proposal, there is another insight as to the mass of the BPS state that both enforces the proposed duality and digs out an interesting feature of the field theory.

It is simpler to explore things heuristically here. While coincident branes give massless modes, a superstring suspended between two distanced D2 or D6-branes gives a BPS hypermultiplet (in our kind of theory, presumably of $\mathcal{N}=(4,4)$ supersymmetry) of mass $\sqrt{\abs{\vec{x}}}$, where $\vec{x}$ is the spatial vector connecting the branes. While a hypermultiplet is massless, a mass is obtained by its coupling to a vector superfield, since the latter obtains a VEV through a Fayet-Iliopoulos $D$-term lying on the U$(1)$ gauge theory in the brane worldvolume. That is, as seen from (\ref{02vectorAction}) and (\ref{02chiralAction}), for a U$(1)$ vector superfield we have a $D$-related action

\begin{equation}
S_D\;=\;\int\frac{1}{g^2} D^2\,+\,\sigma D\bar{\sigma}\,-\,\xi D
\end{equation}
where the last term is the Fayet-Iliopoulos term. After integrating out the auxiliary field $D$, the potential energy $V=g^2(\abs{\sigma}^2-\xi)^2$ is formed which yields the new classical vacuum

\begin{equation}
\langle\sigma\rangle\;=\;\sqrt{\xi}
\end{equation}
which in turn couples to the hypermultiplet and is felt as a mass.

When instead we have two stacks, one of $n_1$ and another of $n_2$ D-branes, we acquire $n_1n_2$ hypermultiplets that transform under the $(n_1,\bar{n}_2)$ representation of U$(n_1)\times$U$(n_2)$. In Hanany-Witten set-ups we have parallel stacks of branes distanced and bordered by NS fivebranes, where the gauge group actually breaks down to SU$(n_i)\times$U$(1)$; the non-trivial U$(1)$ center provides a Fayet-Iliopoulos $D$-term whose coupling is identified with $\xi=\abs{\vec{x}}$. That is, the $D$-term coupling is given by the distances between the NS fivebranes \cite{Hanany:1996ie,Hanany:1997gh}
\begin{equation}
\xi\;=\;\rho_{i+1}-\rho_i
\end{equation}\\
Each U$(1)$ is actually the center of mass of the stack of branes and $D$ is really its Hamiltonian function, where the Fayet-Iliopoulos coupling reflects the fact that we may always add a constant to such a function. While this story is generally studied, let us bring it down onto our case and clarify how it actually works.

By adding a Fayet-Iliopoulos D-term to the $\mathcal{N}=(4,4)$ vector superfield action and integrating out $D$, we acquire the new vacuum $\langle\sigma_i\rangle=\sqrt{\rho_{i+1}-\rho_i}=1/2$. As restated, $\sigma_i$ is one of the scalars of the $\mathcal{N}=(0,4)$ twisted hypermultiplet inside the vector superfield on a stack of D2 or D6-branes, placed between the $(i+1)$th and $i$th stack of NS fivebranes. Notice here that we also normalized, by a redefinition, the fundamental $\rho$-interval distance $\rho_{i+1}-\rho_i=2\pi$ to $1/4$, for convenience that will become apparent momentarily. Now, this VEV gives a mass to a $\mathcal{N}=(4,4)$ hypermultiplet coupled to it and, in particular for our operator of interest, this is achieved through the interactive terms (\ref{MesonEterm}) and (\ref{MesonJterm}) that we brought up in the previous section. That is, if we choose to consider the $\sigma_i$ scalar inside the vector superfield and the $\phi_i$ scalar inside the hypermultiplet then a mass is acquired by the latter as
\begin{equation}
\abs{E_{\Gamma_i}}^2\;=\;\langle\sigma_i\rangle^2\,\abs{\phi_i}^2\;=\;\frac{1}{4}\abs{\phi_i}^2
\end{equation}\\
Accordingly, for other choices of scalar fields inside those multiplets the mass is obtained through other $E$-terms or superpotential $\abs{J}^2$ terms with $J$ as in (\ref{MesonJterm}).

Now, each such hypermultiplet is actually linked to two stacks of D-branes (gauge nodes), one on its left and one on its right along the $\rho$ dimension. This means that the mass that is gained comes from two VEV contributions, that is
\begin{equation}
\abs{E_{\Gamma_i}}^2\,+\,\abs{E_{\Gamma_{i+1}}}^2\;=\;\left(\langle\sigma_i\rangle^2\,+\,\langle\sigma_{i+1}\rangle^2\right)\,\abs{\phi_i}^2\;=\;\frac{1}{2}\abs{\phi_i}^2
\end{equation}\\
where the mass is now unity. Notice that the value of the mass comes from normalization and thus it is a matter of convention on  absolute distances along the $\rho$-dimension. What really matters though is the relative positions of NS fivebranes; changing those shifts the masses of the hypermultiplets in between. Since all the NS fivebranes in our brane set-up are equally separated along $\rho$, accordingly all masses will be the same. Moreover, note that there are as many massive hypermultiplets as the U$(1)$'s. That is, all hypermultiplets between the gauge nodes along the quiver chain are massive. Therefore we only care about the number of those hypermultiplets that contribute to our operator.

Ultimately, the meson operator (\ref{MesonOperator}) contains $m-k$ scalar fields $\phi_i$ which are massive, \textit{associating} the operator itself with a total classical mass

\begin{equation}
M_{\mathcal{M}}\;=\;m-k
\end{equation}
which exactly agrees with the mass (\ref{BPSmass}) of the BPS string.

In regard to our particular choice of the BPS operator, besides the agreement on the dual masses it is worth emphasizing the way that this equality is supported. That is, as with the R-charge (or even the presumable anomalous dimension), it again takes both scalar fields $\phi_i$ and $\sigma_i$ to holographically reflect a dual semiclassical soliton; the $\sigma_i$'s adjust a mass (and a R-charge) and the $\phi_i$'s realize it.

Again, it is the UV particle theory that shapes the proposed meson operator $\mathcal{M}$ and not the actual IR SCFT that sits on the dual side of our AdS$_3$ supergravity backgrounds. While this cautions us to be careful about our statements on what the actual dual BPS operator looks like, we are encouraged by the agreement in mass to make an otherwise bold conjecture: if the choice of operator is correct, then the operator mass somehow \textit{transforms} into a scaling dimension. This is not as presumptuous as it may sound if we consider that the non-perturbative anomalous dimension $\Delta_{\mathcal{M}}=\gamma(g)=m-k$, that we expect, should be generated by the same interactions that produced the Fayet-Iliopoulos mass. Thus the aforementioned transformation is really thought to be a change on how we realize the same field interactions at different energy scales. That is, the interactions given by (\ref{MesonEterm}) and (\ref{MesonJterm}) may be realized as a classical mass in the UV or an anomalous dimension in the IR. This idea is strongly advocated by the fact that the coupling is relevant at the IR of the two-dimensional quantum theory, where the quantum corrections should be important and the scalar masses get integrated out.

As a final comment, the BPS string is a semiclassical bound state which inspires us to assume that its dual operator should too reflect a bound state of two-dimensional fields. That being said, we notice that the operator mass is a sum of all the individual scalar field masses, a fact which renders the UV operator indeed very much alike to a classical bound state of particles. This is a statement on classical bound states in the sense that we neglect an unimportant interaction energy, as we already did with the implicit quantum corrections between fields inside the operator or with the sphere fluctuations on the string mass. While the latter is geometrically obvious through (\ref{BPSmass}), the former may be supported by the fact that the gauge coupling is irrelevant at the UV of two-dimensional quantum field theory.

\subsection{An alternative operator}
Although the last two sections follow the standard examples in the literature (e.g. see \cite{Bergman:2020bvi}), there is an alternative choice of bosonic operator dual to the suspended string. Such an operator may be built out of spinor products, which render it bosonic, as long as it satisfies the desired holographic features, i.e. the correct conformal dimension and R-charge.

This can be achieved through products of left and right-handed spinors inside the $\mathcal{N}=(4,4)$ hypermultiplets that connect the two flavor nodes at stake. Ultimately, the operator reads

\begin{equation}
\mathcal{M}_{k,m}\;=\;\bar{\chi}_+^{(k)}\cdot\bar{\chi}_-^{(k)}\left(\prod_{i=k}^{m-1}\bar{\psi}_+^{(i)}\cdot\bar{\lambda}_-^{(i)}\right)\bar{\chi}_+^{(m)}\cdot\bar{\chi}_-^{(m)}\label{SpinorOperator}
\end{equation}\\
where $\chi_\pm$, $\psi_+$ and $\lambda_-$ are chiral spinors inside the $(4,4)$ hypermultiplets. Again, $\chi_\pm$ are spinors inside the end-point hypermultiplets connecting to the flavor nodes. The operator transforms in the bi-fundamental representation of SU$(N_k)$ $\times$ SU$(N_m)$ and comprises of mass dimension $\Delta_{\mathcal{M}}^0=m-k$ (since $[\psi]=m^{\frac{1}{2}}$ in two dimensions) and R$[\mathcal{M}]=m-k$, since R$[\psi_+]=-1$ and R$[\lambda_-]=0$. Both of those features are exactly what we need.

Though unusual, the new UV operator constitutes a good holographic fit for the suspended string; maybe, it is even better than the more conventional choice of the previous sections, considering that we do not have to assume an IR anomalous dimension or anything else. Nonetheless, there is no obvious reason to choose between the given options of dual operators; as long as the IR SCFT is in the shadows, both of them could be correct. In fact, we could also build operators that are combinations of those two, which would also fit the desired standards. As a final remark, note that even if the scaling dimension of the operator (\ref{SpinorOperator}) exhibits small corrections in the IR, this holographically agrees with the small mass corrections of the BPS string due to its S$^2$-fluctuations that we neglected in (\ref{BPSmass}).\\

\section{Epilogue}
Summarizing, in Section \ref{Section3} we studied all possible categories of vacua within a particular AdS$_3$ family of massive IIA supergravity solutions, first given in \cite{Lozano:2019ywa}. Apart from the original solutions introduced there, we presented the remaining types of vacua in the same family which all naively seem to give anomalous dual quiver gauge theories. We proved that these erratic solutions imply D-branes on the boundary of the space, which in turn correspond to flavor symmetries that exactly cancel the apparent gauge anomalies. A special feature of the situation is that, due to strong RR fluxes on the boundary of the space, these D-branes come exclusively in bound states forming polarizations that provide the quiver with flavor in a quite idiosyncratic way.

After dealing with all possible kinds of solutions and quiver theories, in Section \ref{Section4} we supplement the quiver structure with additional matter in the form of bifundamental links between color and flavor nodes. These, we argue, may only be $\mathcal{N}=(4,4)$ hypermultiplets corresponding to suspended superstrings between D2 and D4-branes or D6 and D8-branes in the ancestral Hanany-Witten set-up.

Having introduced the complementary bifundamental matter too, in Section \ref{Section5} we put holography to the test by considering a semiclassical string inside the AdS$_3$ background stretched between two D-branes. We call this a meson string and by finding its mass and R-charge we show it is a BPS state. Next, we propose a UV operator dual to the soliton and we argue that there is a unique choice of fundamental scalar fields that synthesize it. Moreover, crucial to the construction of this operator is the additional bifundamental matter we have introduced. While the R-charge of the proposed operator seems to get along with our expectations, its conformal dimension is classically zero since scalar fields in two spacetime dimensions have a vanishing mass dimension. What is more, since the two-dimensional SCFT we are assuming is strongly coupled and these AdS$_3$ vacua have been proven to be non-integrable, the perturbative regime of calculations is out of our reach. Nonetheless, by bringing to the surface the superpotential of the UV quiver theory, we find interactions between the scalars inside the operator and we are led to the conclusion that the latter should acquire a totally non-perturbative anomalous dimension at the IR, equal to the mass of the BPS string.

Pursuing the holographic picture of the meson string, we focus on the quiver structure and find that scalars inside the vector superfields should obtain a VEV through a Fayet-Iliopoulos term. The latter is due to the U$(1)$ theory inside the U$(N)$ gauge group of each stack of branes in the set-up. Superpotential interactions between the vector and hypermultiplets then dictate that bifundamental matter acquires a mass, ultimately associating the dual meson operator with a classical mass equal to that of the BPS string. Since the operator mass is a sum of all the individual scalar field masses, this renders the operator indeed very much alike to a classical bound state of particles dual to a bound string state between D-branes.\\

\paragraph*{Acknowledgments}
I thank C. Nunez for his help on this work, along with Y. Lozano, N. Macpherson and A. Tomasiello for our useful correspondence. I also thank S. Speziali, D. Thompson and D. Giataganas for their comments and an anonymous reviewer of an earlier version of the manuscript for pointing out a crucial mistake in the calculations. This work is supported by both a STFC scholarship and the Swansea University College of Science.\newpage

\appendix

\section{Extremal $p$-brane solutions}\label{appendixA}
Extremal $p$-branes are supergravity solutions that in the context of superstring theory are identified with stacks of D$p$-branes. These are distinct from O-planes that essentially constitute boundary conditions for strings. The leading order backgrounds for all the above read

\begin{equation}
\begin{split}
\mbox{$p$-brane}\hspace{0.56cm}:&\hspace{0.5cm}\dd s^2\sim x^{\frac{7-p}{2}}\dd s^2_{\tiny\mbox{M$^{1,p}$}}+x^{\frac{p-7}{2}}\left(\dd x^2+x^2\dd s^2_{\tiny\mbox{$\Sigma^{8-p}$}}\right)\,,\hspace{2cm}e^\phi\sim x^{\frac{(3-p)(p-7)}{4}}\,,\\[10pt]
  \begin{array}{c}
   \text{$p$-brane} \\ 
    \text{smeared on $\tilde{\Sigma^s}$}
  \end{array}:&\hspace{0.5cm}\dd s^2\sim x^{\frac{7-p-s}{2}}\dd s^2_{\tiny\mbox{M$^{1,p}$}}+x^{\frac{p+s-7}{2}}\left(\dd x^2+\dd s^2_{\tiny\mbox{$\tilde{\Sigma}^s$}}+x^2\dd s^2_{\tiny\mbox{$\Sigma^{8-p-s}$}}\right)\,,\hspace{0.2cm}e^\phi\sim x^{\frac{(3-p)(p+s-7)}{4}}\,,\\[10pt]
  \mbox{O$p$-plane}\hspace{0.56cm}:&\hspace{0.5cm}\dd s^2\sim\frac{1}{\sqrt{x}}\dd s^2_{\tiny\mbox{M$^{1,p}$}}+\sqrt{x}\left(\dd x^2+\dd s^2_{\tiny\mbox{$\Sigma^{8-p}$}}\right)\,,\hspace{3cm}e^\phi\sim x^{\frac{3-p}{4}}\,,
\end{split}
\end{equation}\\
where we schematically acknowledge constants. Here M$^{1,p}$ is a manifold that the brane fills, $\Sigma^{8-p}$ is a compact space $-$ on which one integrates to obtain the associated charge of the brane $-$ and $\tilde{\Sigma}^s$ is the manifold over which a brane may be smeared.\\

\section{Two dimensional $\mathcal{N}=(0,4)$ superfields}\label{appendixB}

\subsection{Field content and action}\label{appendixFields}
Traditionally, extended supersymmetric theories are best realized through constituent, minimal supersymmetric multiplets. $\mathcal{N}=(0,4)$ supersymmetry is no different and boils down to $\mathcal{N}=(0,2)$ superfields, which we now introduce. The language and content we present is mainly based on \cite{Witten:1993yc,Tong:2014yna}, which both hold excellent reviews on the subject.

\paragraph{Gauge multiplet}
This is a real superfield, $\mathcal{V}$, which comprises of an adjoint-valued complex left-handed fermion $\zeta_-$, a real auxiliary field $D$ and a gauge field $A$. The standard kinetic term for the gauge multiplet expands into the action

\begin{equation}
S_{\mbox{\tiny gauge}}=\frac{1}{g^2}\Tr\int\dd^2x\left(\frac{1}{2}F_{01}+i\bar{\zeta}_-(\mathcal{D}_1+\mathcal{D}_1)\zeta_-+D^2\right)\,.\label{02vectorAction}
\end{equation}

\paragraph{Chiral multiplet}
A $\mathcal{N}=(0,2)$ chiral superfield, $\Phi$, comprises of a right-moving fermion $\psi_+$ and a complex scalar $\phi$, which both transform in the same gauge group representation. The kinetic term for the gauged chiral multiplet expands into

\begin{equation}
S_{\mbox{\tiny chiral}}=\int\dd^2x\left(-\abs{D_\mu\phi}^2+i\bar{\psi}_+\left(D_0-D_1\right)\psi_+-i\bar{\phi}\zeta_-\psi_++i\bar{\psi}_+\bar{\zeta}_-\phi+\bar{\phi}D\phi\right)\,.\label{02chiralAction}
\end{equation}

\paragraph{Fermi multiplet}
This is an anticommuting superfield, $\Psi$, containing a left-moving spinor $\psi_-$ and a complex auxiliary field $G$. The Fermi superfield is constrained by $\bar{\mathcal{D}}_+\Psi=E$ where $\mathcal{D}_+=\partial_{\theta^+}-i\bar{\theta}^+(\mathcal{D}_0+\mathcal{D}_1)$, with $\mathcal{D}_{0,1}=\partial_{0,1}+iA_{0,1}$ and $E=E(\Phi_i)$ a holomorphic function of the chiral superfields $\Phi_i$. The kinetic term for the Fermi multiplet expands into

\begin{equation}
S_{\mbox{\tiny Fermi}}=\int\dd^2x\,\left(i\bar{\psi}_-(D_0+D_1)\psi_-+\abs{G}^2-\abs{E(\phi_i)}^2-\bar{\psi}_-\pdv{E}{\phi_i}\psi_{+i}+\bar{\psi}_{+i}\pdv{\bar{E}}{\bar{\phi}_i}\psi_-\right)\,.\label{FERMImulti}
\end{equation}\\
The holomorphic function $E(\phi_i)$ comes up as a potential $\sim\abs{E(\phi_i)}^2$ inside the action and thus its particular choice, along with superpotential terms, determine the interactions of the theory.\\

\paragraph{Superpotentials}
Considering multiple Fermi superfields $\Psi_a$ which couple to scalar chiral superfields $J^a(\Phi_i)$ through $S_J\sim\int\Psi_aJ^a$ over half of the superspace, supersymmetry dictates that superfields are constrained as $E\cdot J=\sum_aE_aJ^a=0$. $J^a(\phi)$ produce potential terms $\sim\abs{J^a(\phi_i)}^2$ which are usually referred to as the \textit{superpotential} in $\mathcal{N}=(0,2)$ theories. Therefore, besides the $E$-terms, the $J$-terms also give potential terms in $\mathcal{N}=(0,2)$ supersymmetric theories, all of them directly connected to Fermi multiplets. The attachment $E\cdot J=0$ when multiple Fermi and chiral multiplets are present, decides for the particular interactions in the theory. But to see how this plays out we must first introduce $\mathcal{N}=(0,4)$ supersymmetric multiplets.\\

Two dimensional $\mathcal{N}=(0,4)$ supersymmetry has four real right-moving supercharges that rotate in the $(\mathbf{2},\mathbf{2})_+$ representation of a SO$(4)_R$ $\cong$ SU$(2)_R$ $\times$ SU$(2)_R$ R-symmetry, where the plus sign indicates the chirality under the SO$(1,1)$ Lorentz group. The superfields in this kind of theories are the following.\\

\paragraph{$\mathcal{N}=(0,4)$ vector multiplet}
Since in two dimensions the gauge field is not propagating it is natural that two-dimensional $\mathcal{N}=(0,4)$ vector superfields are composed of left-handed spinors, which don't transform under right-moving supersymmetry. Thus, a $\mathcal{N}=(0,4)$ vector superfield consists of an adjoint-valued $\mathcal{N}=(0,2)$ Fermi superfield $\Theta$ and a $\mathcal{N}=(0,2)$ vector superfield .

Besides the gauge field, there are two left-handed complex fermions, $\zeta_-^a$ and three auxiliary fields, transforming in the $\mathbf{(2,2)_-}$ and $\mathbf{(3,1)}$ R-symmetry representations, respectively. The Fermi superfield is constrained through $\bar{\mathcal{D}}_+\Theta\;=\;E_\Theta$ with $E_\Theta$ depending on the matter content, i.e. the chiral superfields present in the theory.\\

\paragraph{$\mathcal{N}=(0,4)$ hypermultiplet}
The first way to couple matter fields to a $\mathcal{N}=(0,4)$ vector multiplet (essentially to its constituent $\mathcal{N}=(0,2)$ Fermi multiplet) is to consider a $\mathcal{N}=(0,4)$ hypermultiplet that consists of two $\mathcal{N}=(0,2)$ chiral superfields, $\Phi$ and $\tilde{\Phi}$, which transform in conjugate gauge group representations and whose pairs of complex scalars and right-handed spinors transform in the $\mathbf{(2,1)}$ and $\mathbf{(1,2)_+}$ representations, respectively, under the R-symmetry.\\

\paragraph{$\mathcal{N}=(0,4)$ twisted hypermultiplet}
Another possible way to couple matter fields to a $\mathcal{N}=(0,4)$ vector multiplet $\mathcal{N}=(0,4)$ is through a twisted hypermultiplet. This consists of a pair of $\mathcal{N}=(0,2)$ chiral multiplets, $\Sigma$ and $\tilde{\Sigma}$, which too transform in conjugate gauge group representations. Now, nonetheless, different R-charge is being enforced by the coupling to the Fermi field $\Theta$. In contrast to hypermultiplets, the scalars and right-handed spinors now transform in the $\mathbf{(1,2)}$ and $\mathbf{(2, 1)_+}$ representations of R-symmetry.\\

\paragraph{$\mathcal{N}=(0,4)$ Fermi multiplet}
Those contain two $\mathcal{N}=(0,2)$ Fermi superfields, $\Gamma$ and $\tilde{\Gamma}$, which transform in conjugate gauge group representations and whose left-moving spinors transform in the $\mathbf{(1,1)_-}$ R-symmetry representation.\\

\paragraph{$\mathcal{N}=(0,2)$ Fermi multiplet}
Finally, it is acceptable in $\mathcal{N}=(0,4)$ supersymmetric theories to consider $\mathcal{N}=(0,2)$ Fermi multiplets, as long as their left-moving spinors are SO$(4)_R$ singlets and, according to that R-symmetry transformation, couple appropriately to the rest of the matter in the theory.\\

As we are about to see, our quantum field theory also contains $\mathcal{N}=(4,4)$ superfields that decompose under $\mathcal{N}=(0,4)$ supersymmetry into their $\mathcal{N}=(0,4)$ superfield constituents. The $\mathcal{N}=(4,4)$ vector multiplet splits into an $\mathcal{N}=(0,4)$ vector multiplet and an adjoint-valued $\mathcal{N}=(0,4)$ twisted hypermultiplet. The chiral superfields $\Sigma$ and $\tilde{\Sigma}$ inside the twisted hypermultiplet couple to the Fermi multiplet $\Theta$ inside the $\mathcal{N}=(0,4)$ vector superfield. Finally, a $\mathcal{N}=(4,4)$ hypermultiplet decomposes into an $\mathcal{N}=(0,4)$ hypermultiplet, $\Phi$ and $\tilde{\Phi}$, and an $\mathcal{N}=(0,4)$ Fermi multiplet, $\Gamma$ and $\tilde{\Gamma}$.\\

\subsection{U(1) R-charge}\label{appendixRanomaly}
From the SU$(2)_R$ $\times$ SU$(2)_R$ R-symmetry of the $\mathcal{N}=(0,4)$ theory, we single out a U$(1)_R$ inside one SU$(2)_R$ and give the U$(1)_R$ charge of each fermion in the above multiplets.\\

For the $\mathcal{N}=(0,4)$ vector multiplet we have that the left-handed fermion inside the vector has R$[\zeta_-]=+1$ while the same holds for the left-handed fermion inside the Fermi multiplet, i.e. R$[\psi_-]=+1$. On the contrary, both right-handed fermions inside the $\mathcal{N}=(0,4)$ twisted hypermultiplet have R$[\psi_+]=0$. For both right-handed fermions inside the $\mathcal{N}=(0,4)$ hypermultiplet we have R$[\psi_+]=-1$. Finally, the fermion inside the $\mathcal{N}=(0,2)$ Fermi multiplet is uncharged under R-symmetry.

\section{The D8/D4 bound state}\label{AppendixD8/D4}
We consider the background of the case with a constant $u$ function and study the beginning of its $\rho$-dimension where D4-branes seem to polarize into a D8/D4 bound state. The fact that $C_9^{el}$ field becomes infinitely strong at that endpoint reasonably makes the D8/D4 bound state dominant, yet a more formal proof of it being the true vacuum is in order.

Comparing to Myers's original calculation \cite{Myers:1999ps}, here we are dealing with higher dimensional branes. Furthemore, the method developed in \cite{Myers:1999ps} holds in the flat space limit, whereas our bound state takes place in AdS$_3\times$S$^2\times$CY$_2\times$I$_\rho$. What is more, Calabi-Yau manifolds lack a particular metric tensor whatsoever.

However, the situation is less dramatic than it may look. First of, the Chern-Simons term

\begin{equation}
S_{\mbox{\tiny CS}}^{\mbox{\tiny D4}}\;=\;\mu_4\int\Tr\:\sum e^{i\lambda\imath_\Phi\imath_\Phi} C_{(n)}e^{\mathcal{F}_2}
\end{equation}\\
gets only deformed away from the flat space limit by terms coupled to the $B_2$ field. These terms would be unimportant compared to our infinitely strong $C_9^{el}$ potential coupling, but the Kalb-Ramond field vanishes at $\rho=0$ for constant $u(\rho)$ anyway. Next, the Dirac-Born-Infeld (DBI) action

\begin{equation}
S_{\mbox{\tiny DBI}}^{\mbox{\tiny D4}}\;=\;-T_4\int\dd^4\xi\,\Tr\:e^{-\phi}\sqrt{-\det(G_{ab}+G_{ai}(Q^{-1}-\delta)^{ij}G_{jb}+\lambda\tilde{f}_{ab})\det(Q^i_j)}
\end{equation}\\
where
\begin{equation}
Q^i_j\;=\;\delta^i_j\,+\,i\lambda[\Phi^i,\Phi^k]G_{kj}\,.
\end{equation}\\
The $a,b$ are indices pulled-back on the D4-brane worldvolumes, while $i,j$ are their transverse dimensions. That is, $G_{\mu\nu}=(G_{ab},G_{ij})$ where $G_{ai}=0$ and the transverse field $G_{ij}$ includes the $\rho$-dimension and an independent CY$_2$ block.

Choosing a static gauge where the D4-branes' worldvolumes fill up AdS$_3\times$S$^2$, i.e. choosing worldvolume coordinates and the transverse modes (which are scalars in the D4 worldvolume) as

\begin{equation}
\xi^a\;=\;X^a\;=\;(t,x,r,\theta,\phi)\;,\hspace{2cm}X^i(\xi^a)\;=\;\lambda\Phi^i(\xi^a)\,,\label{staticgauge}
\end{equation}\\
where the $\lambda$ was included on dimensional grounds, then the DBI action reads
\begin{equation}
S_{\mbox{\tiny DBI}}^{\mbox{\tiny D4}}\;=\;-T_4\int\dd^4\xi\,\Tr\:e^{-\phi}\sqrt{-\det(G_{ab}+\lambda^2\partial_a\Phi^i\partial_b\Phi^jG_{ij})\det(\delta^i_j\,+\,i\lambda[\Phi^i,\Phi^k]G_{kj})}
\end{equation}\\
where we ignored the D4-brane gauge field $\tilde{f}$ as unimportant. Using the fact that the determinant behaves like $\det(A+\lambda B)=\det A+\lambda\Tr B+\ldots$ for small $\lambda$, we obtain the potential energy

\begin{equation}
V(\Phi)\;=\;N_4T_4M_4\,-\,\frac{T_4M_4\lambda^2}{4}\Tr\,[\Phi^i,\Phi^j]^2\,-\,i\frac{T_4M_4\lambda^3}{12}\Tr\,[\Phi^i,\Phi^j]^3\,+\,\ldots\label{SYMpot}
\end{equation}\\
where the ellipsis contains higher-order potential terms and contractions with the transverse metric $G_{ij}$ are implied. $N_4$ is the number of D4-branes and $M_4$ comes from the factor $e^{-\phi}\det G$, which for our background (\ref{exampleIIu=cD8}) at $\rho\rightarrow0$ scales as

\begin{equation}
e^{-\phi}\det G\;\;\xrightarrow{\rho\rightarrow0}\;\;M_4\,\rho^{\frac{5}{4}}\rho^{-\frac{5}{4}}\;=\;M_4
\end{equation}\\
which goes to a constant. Notice that in the flat space limit, the second term of (\ref{SYMpot}) reflects the familiar supersymmetric Yang-Mills (SYM) potential.

So far, the sole deviation from the flat space analysis is the contraction of indices in the potential (\ref{SYMpot}) with the transverse metric $G_{ij}$. This field includes the $\rho$-dimension component and an independent CY$_2$ block. The former is known but unimportant since the $\Phi^\rho$ modes will not be ultimately involved in the potential energy and thus no such indices will need to contract, while the latter is essential but lacks a particular metric tensor. We could maybe realize some generic algebraic constraints on the Calabi-Yau block, like its Ricci flatness, but we do need a particular metric tensor which makes it is easier to assume CY$_2=$ T$^4$ and thus let for a Euclidean $\mathbb{R}^4$ metric.

Our study significantly simplifies by choosing a convenient gauge for the RR potential as

\begin{equation}
C_9^{el}\;=\;-\frac{u^2}{h_8}\mbox{vol(AdS$_3$)}\wedge\mbox{vol(S$^2$)}\wedge\mbox{vol(CY$_2$)}\,.
\end{equation}\\
On these grounds, while picking the static gauge (\ref{staticgauge}), we can expand the source term

\begin{equation}
\begin{split}
S_{\mbox{\tiny CS}}^{\mbox{\tiny D8/D4}}\;=\;-\frac{\lambda^2}{2}\mu_8\int\Tr\:(\imath_\Phi\imath_\Phi)^2\,C_9^{el}\;&=\;-\frac{\lambda^2}{2}\mu_4\int\dd^5\xi\,\Tr\:\Phi^i\Phi^j\Phi^k\Phi^l\,C_{ijkltxr\theta\phi}\\
&=\;-\frac{\lambda^2}{8}\mu_4\int\dd^5\xi\,\Tr\:[\Phi^i,\Phi^j][\Phi^k,\Phi^l]\,C_9
\end{split}\label{AppendixD8/D4WZterm}
\end{equation}\\
where we redefine the Latin letters $i,j,k,l$ to denote only CY$_2$ directions. The transverse modes $\Phi^i$ are in general anticommuting matrices, where the diagonal elements are the positions of the D4-branes, while the non-diagonal ones reflect their quantum geometry due to the superposition of strings ending on them. The fact that $\Phi^i$ are oscillations in non-flat dimensions is not restrictive in any way, since we fundamentally assume those modes as generic anticommuting matrices that may (and actually do) give a fuzzy geometry. Also, note that in general we should include $\Phi^\rho$ too, but not in our particular gauge of $C_9^{el}$.

Now we want to focus on $\rho=0$ where all the action takes place, i.e. expand $C_9^{el}$ around that endpoint. It being a singular endpoint implies a Laurent expansion but, since it is also the endpoint of a closed interval, this series is not well defined around it. Thus, we just pick a point $x$ \textit{close} to $\rho=0$ and expand around it, inside a circular region (of the complex domain) $-$ of radius $x$ too $-$ which touches the singularity. That is, the expansion reduces to a Taylor series around $x$ as

\begin{equation}
S_{\mbox{\tiny CS}}^{\mbox{\tiny D8/D4}}\;=\;-\frac{\lambda^2}{8}\mu_4\int\dd^5\xi\,\Tr\:[\Phi^i,\Phi^j][\Phi^k,\Phi^l]\,\Big(C_9|_{\rho=x}+\lambda\Phi^\rho F_{10}|_{\rho=x}+\ldots\Big)\,.
\end{equation}\\
Since $h_8\rightarrow0$ for small $x$, the RR fields $C_9$ and $F_{10}$ blow up there and thus from now on we will consider them as largely valued quantities.

The above source term adds to the interactions (\ref{SYMpot}) of the DBI action and hence, taking into account the full D4-brane action $S=S_{\mbox{\tiny DBI}}+S_{\mbox{\tiny CS}}$, we acquire the potential energy

\begin{equation}
\begin{split}
V(\Phi)\;=\;&-\,\frac{\lambda^2}{4}\Tr\,[\Phi^i,\Phi^j]^2\,+\,\frac{\lambda^2}{8}\Tr\,[\Phi^i,\Phi^j][\Phi^k,\Phi^l]C_9|_{\rho=x}\\
&-\,i\frac{\lambda^3}{12}\Tr\,[\Phi^i,\Phi^j]^3\,+\,\frac{\lambda^3}{8}\Tr\,[\Phi^i,\Phi^j][\Phi^k,\Phi^l]\Phi^\rho\,F_{10}|_{\rho=x}
\end{split}
\end{equation}\\
where we have assumed a constant mode $\Phi^\rho$ to simplify the game and reparametrized the fields conveniently to absorb numerical factors. Reparametrizing once more, the potential gets an order by order variation $\pdv{V}{\Phi}=0$ as

\begin{equation}
\begin{split}
\mathcal{O}(\lambda^2)\,:\hspace{1.5cm}&[\Phi^i,\Phi^j]\;=\;[\Phi^k,\Phi^l]\,C_{ijkl...}\\[10pt]
\mathcal{O}(\lambda^3)\,:\hspace{1.5cm}&[\Phi^i,\Phi^j][\Phi^j,\Phi^k]\;=\;-i[\Phi^l,\Phi^m]F_{iklm...}
\end{split}
\end{equation}\\
which has a trivial solution $[\Phi^i,\Phi^j]=0$ giving $V_0=0$, corresponding to separated D4-branes. Alternatively, combining both of these equations, the potential also exhibits the non-trivial solution

\begin{equation}
[\Phi^i,\Phi^j]\;=\;-i\epsilon^{ij}\partial_\rho
\end{equation}\\
which in momentum space reads

\begin{equation}
[\Phi^i,\Phi^j]\;=\;\epsilon^{ij} p_\rho\label{AppendixNontrivialSol}
\end{equation}\\
where we abuse the antisymmetric tensor just to sustain the antisymmetry of the commutator into the rhs. Placing this solution back into the SYM potential we get

\begin{equation}
V_\star\;\cong\;\lambda^2\,p_\rho^2\,C_9|_{\rho\rightarrow0}\,+\,\mathcal{O}(\lambda^3)
\end{equation}\\
where we used the fact that $C_9$ is large at $\rho\rightarrow0$.

As a matter of fact, $C_9$ is not only large but also negative at that endpoint, which means that $V_\star<0$. Since the separated D4-branes correspond to the null energy state $V_0=0$, the latter is unstable and condenses out into the non-trivial D8/D4 bound state with $V_\star$ which is the true stable vacuum at $\rho=0$. Also, notice the fact that specifically $V_\star\rightarrow-\infty$, due to the strong RR potential $C_9\rightarrow-\infty$ at $\rho\rightarrow0$, which saves us from having to also investigate other bound states. In our case, $C_3^{el},C_7^{el}\rightarrow0$ at $\rho\rightarrow0$ anyway, but even if this was not the case there just cannot be any lower energy than $V_\star$.\\

\section{R-charge of the BPS state}\label{appendixRcharge}
Naively, the $B_2$ field in (\ref{GeneralAdS3background}) has nothing to do with the 1-form $\cos\theta\,\dd\phi$. However, $B_2$ exhibits large gauge transformations across the $\rho$-intervals $[2\pi k,2\pi(k+1)]$, which are explicitly realized through the 1-form

\begin{equation}
\Lambda_1=\Theta\big(\rho-2\pi k\big)\,\Theta\big(2\pi(k+1)-\rho\big)\pi k\cos\theta\,\dd\phi\,.
\end{equation}\\
Therefore,  the large gauge transformations $B_2\rightarrow B_2+\dd\Lambda_1$ read

\begin{equation}
\begin{split}
B_2\longrightarrow B_2\;+\;\Theta\big(\rho-2\pi k\big)&\,\Theta\big(2\pi(k+1)-\rho\big)\,\pi k\,\dd\Omega_2\\
&+\;\left[\delta\big(\rho-2\pi k\big)-\delta\big(2\pi(k+1)-\rho\big)\right]\,\pi k\,\dd\rho\wedge\cos\theta\,\dd\phi
\end{split}\label{B2largeGaugeTrans}
\end{equation}\\
where, in this explicit formulation, the only difference now is the novel delta-terms, $B_2^\delta$. The latter, which are the ones producing the R-charge, are integrated over a $\rho$-interval as

\begin{equation}
\begin{split}
\frac{1}{2\pi}\int B_2^\delta\;=\;\frac{2}{2\pi}\int_{\mathbb{R}}\cos\theta\,\dd\phi\int_{2\pi k}^{2\pi(k+1)}\dd\rho\,\bigg\lbrace&\left[\delta\big(\rho-2\pi k\big)-\delta\big(2\pi(k+1)-\rho\big)\right]\pi k\\
&-\delta\big(2\pi k-\rho\big)\pi(k-1)+\delta\big(\rho-2\pi(k+1)\big)\pi(k+1)\bigg\rbrace
\end{split}
\end{equation}\\
where the first line is the contribution coming from $B_2^\delta$ defined on the interval $[2\pi k,2\pi(k+1)]$ as expected, while the second line includes the contributions coming from the intervals prior and next to that. Considering $\int_0^\infty\delta(x)\dd x=1/2$, the above integral gives

\begin{equation}
\frac{1}{2\pi}\int B_2^\delta\;=\;\int_{\mathbb{R}}\cos\theta\,\dd\phi
\end{equation}\\
and the whole meson string $\mathcal{M}_{k,m}$ acquires the R-charge source term

\begin{equation}
S_{\mbox{\tiny$\mathcal{M}$}}\;=\;\left(m-k\right)\int_{\mathbb{R}}\cos\theta\,\dd\phi
\end{equation}\\
which yields its R-charge

\begin{equation}
Q_R\;=\;m-k\,.
\end{equation}\\

\end{document}